\documentclass[reprint,aps,pre,groupedaddress]{revtex4-1}
\usepackage{amsmath,amssymb}
\usepackage{natbib}
\usepackage{labelfig,rotating}
\usepackage{epsf,graphicx,psfrag}

\newcommand\Rey{\mbox{\textit{Re}}}  
\usepackage{color}

\begin{document}
\title{Localization in a spanwise-extended model of plane Couette flow}
\author{M. Chantry}
\email{Matthew.Chantry@bris.ac.uk}
\author{R. R. Kerswell}
\email{R.R.Kerswell@bris.ac.uk}
\affiliation{School of Mathematics, University of Bristol, Bristol BS8 1TW, UK.}
\date{\today}


\begin{abstract}

We consider a 9-PDE (1-space and 1-time) model of plane Couette flow
in which the degrees of freedom are severely restricted in the
streamwise and cross-stream directions to study spanwise localisation
in detail. Of the many steady Eckhaus (spanwise modulational) instabilities
identified of global steady states, none lead to a localized
state. Localized periodic solutions were found instead which arise
in saddle node bifurcations in the Reynolds number.  These solutions
appear global (domain filling) in narrow (small spanwise) domains yet
can be smoothly continued out to fully spanwise-localised states in
very wide domains. This smooth localisation behaviour, which has also
been seen in fully-resolved duct flow (Okino 2011), indicates that an
apparently global flow structure needn't have to suffer a modulational
instability to localize in wide domains.

\end{abstract}

\pacs{}

\maketitle

\section{Introduction}

The transition to turbulence in wall-bounded shear flows is an old and
intriguing problem in fluid dynamics since transition is typically
observed even though the laminar state is stable. The discovery in the
last twenty years of exact but unstable solutions
(e.g. \cite{nagata90,waleffe98,faisst03,wedin04}) surrounding the
laminar state in phase state has helped explain how transitional flows
can be so complicated. Progress continues to be made using the growing
library of solutions and ideas from dynamical systems theory to
understand the dynamics as a walk in phase space between such
solutions \citep{Tutty2007,gibson08,Kawahara2012,willis12,Kreilos2014}.
However computational limitations have restricted much of this work to
small domains periodic in space where turbulence is global. Extending
this approach to physically-realised domains requires localized solutions.

The first of these - a solution localized in the spanwise direction -
was found via edge tracking by \cite{schneider10localized} in plane
Couette flow (pCf). Perhaps more important was the discovery via
continuation in Reynolds number of the connection of the localized
solution branch to a bifurcation from a spatially-periodic solution
\citep{schneider2010}. This opened up the possibility of using the
catalogue of known spatially-global solutions to find localized
versions. However, edge tracking has continued to be the method of
choice (e.g. in pipe \cite{avila2013,chantry2013}; channel flow
\cite{Zammert2014a,Zammert2014b} and pCf
\cite{Melnikov2013}) with subsequent connections found only later
moving \textit{from} the localized solutions \textit{to} global
solutions \cite{chantry2013}. 

One novel approach to going {\em from} spatially-periodic states {\em
  to} localised states has been to attempt to converge cleverly-masked
versions of global solutions as localised solutions of the governing
equations \cite{gibson2013,Brand2014}. While successful, this has not
informed the important question of how generic bifurcations are which
lead to localized solutions when viewed amongst all the bifurcations
experienced by the spatially periodic solutions. The numerical cost of
finding and tracking bifurcations in the large domains necessary to
see localisation, makes seeking an answer an intimidating
proposition. Nevertheless Melnikov et al. \cite{Melnikov2013} have
made a start by studying long spanwise wavelength instabilities of one
global state in pCf. However, they did not show how such bifurcations
could lead to a localised state although edge tracking indicated
localised states did exist nearby. The motivation for this study was
to introduce a more-accessible low-dimensional model to explore
the result of bifurcations from spanwise-periodic states and the existence of
localised states. A particular objective was to see if the homoclinic
snaking
\citep{champneys1998homoclinic,burke2007homoclinic,chapman2009exponential}
seen in \cite{schneider10localized} could be captured here.

Low-dimensional models have been employed in the past to try to
understand facets of turbulent shear flows with a good track record of
success. Waleffe \cite{waleffe97} constructed a hierarchy of models for pCf
consisting of ordinary differential equations (ODEs) describing the
evolution of prescribed flow modes representing the crucial flow
ingredients observed in experiments. He demonstrated that a
self-sustaining process could transfer energy between the mean flow,
rolls, streaks and wave-like instabilities in a cycle but these simple models 
could not reproduce the complex temporal dynamics of turbulence.
Moehlis et al. \cite{moehlis1} adapted Waleffe's 8-ODE model to include
higher harmonic interactions by adding a further ODE. This 9-ODE model
showed many of the rich temporal dynamics of turbulence with
sensitivity to initial conditions, memoryless decay and exact
solutions \citep{moehlis2}. Dawes and Giles \cite{dawes2011} carried out a different
extension of the 8-ODE model by removing the
prescribed spanwise dependence of the modes leading to 8 PDEs in the
spanwise coordinate, $z$, and time $t$. This extension also generated
a model with many of the characteristics of turbulence but the focus
was on spanwise-limited (narrow) domains. In this work we combine
these two extensions to study a 9-PDE extension of the
9-ODE model over spanwise-extended (wide) domains to examine how the
flow can spanwise-localise. This 9-PDE model,
which restricts the dynamics in 2 spatial directions but fully resolves the
remaining spanwise direction, is similar in spirit to previous reduced
models which fully resolve 2 spatial directions and use a reduced
resolution in the other \cite{Locher2000,Manneville2004,Lagha2007,
  Willis2009}.  If \citep{Willis2009} described their model as a
`2+$\epsilon$' dimensional model, then our model here is `1+2$\epsilon$'
dimensional.


This paper is organized as follows. Section \ref{sec:model} (supported by Appendix
\ref{app:eqn}) describes the derivation of the reduced model which is closely
related to the 8-PDE model of \cite{dawes2011}, and briefly discusses
the spatial Floquet theory employed to identify the Eckhaus
(modulational) instabilities of interest here. Section
\ref{sec:modres} details the spanwise-periodic states found in the
model, the steady modulational instabilities they possess and the
results of tracking all of these bifurcations. Section \ref{sec:local} applies
edge tracking to  find localized solutions and section \ref{sec:walls} explores
the effect of different spanwise boundary conditions. A final section
discusses our results and their implications for fully-resolved models
of the wall-bounded shear flows.


\section{Model derivation}
\label{sec:model}

\begin{figure}
\begin{center}
\SetLabels 
(-0.01*0.05){\normalsize $-1$} \\
(-0.00*0.485){\normalsize 0} \\
(-0.00*0.92){\normalsize $1$} \\
(-0.06*0.485){\normalsize $y$} \\
(-0.02*-0.05){\normalsize $-\sqrt{2}$} \\
(0.5*-0.05){\normalsize $0$} \\
(0.95*-0.05){\normalsize $\sqrt{2}$} \\
(+0.5*-0.13){\normalsize $U(y)$} \\
\endSetLabels 
\leavevmode
\strut\AffixLabels{\includegraphics[width=0.6\columnwidth]{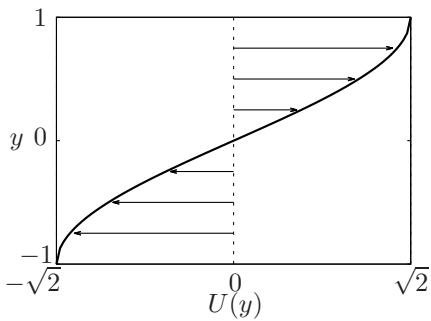}}\\
\vspace{3mm}
\end{center}
\caption{Laminar flow $\mathbf{u}=\sqrt{2} \sin \left(\beta y \right) \mathbf{e_x}$.}

\label{fig:base}
\end{figure}

To study spanwise localization processes, a set of 9 PDEs in $z$ and
$t$ are derived with the various steps closely following
\cite{dawes2011}. Plane Couette flow is the flow between two parallel
plates moving in their plane in opposite directions. The coordinate
system is defined with $x$ in the downstream direction, $y$ in the
shear-dependent direction and $z$ in the spanwise direction. The
Navier-Stokes equation is non-dimensionalized by $h$, the half
distance between the walls and $U_0$ the laminar flow speed at $y=\pm
h/2$ (this differs from pCf where wall speed is used). The Reynolds
number in this system is defined as $\Rey = U_0 h / \nu$, where $\nu$
is the kinematic viscosity. In sections \ref{sec:modres} and \ref{sec:local}, we consider a domain
periodic in $x$ and $z$ with impermeable, stress-free boundary
conditions in $y$ (in section \ref{sec:walls}, the boundary conditions in $x$ are
changed to non-slip).  For $\mathbf{u}=\left(u,v,w\right)$ these
conditions are written as
\begin{align}
\mathbf{u}(x,y,z)=\mathbf{u}(x+L_x,y,z)=\mathbf{u}(x,y,z+L_z)\\
v(x,\pm1,z,t)=\left.\frac{\partial u}{\partial y}\right|_{y=\pm1}
=
\left.\frac{\partial w}{\partial y}\right|_{y=\pm1}=0.
\end{align}
The dimensions of the domain are $\left[0,L_x\right] \times
\left[-1,1\right] \times \left[0,L_z\right]$. The system is driven by
the body force,
$$\mathbf{F}\left(y\right):= \frac{\sqrt{2} \beta^2}{R}\sin
\left(\beta y\right) \mathbf{e_x},$$ where $\beta =
\frac{\pi}{2}$. This results in a steady unidirectional laminar flow
$$\mathbf{u}=\sqrt{2} \sin \left(\beta y \right) \mathbf{e_x}.$$ This
forcing was chosen by Waleffe to produce a base flow close to that of
pCf but using Fourier modes. Despite the inflection point this flow
remains linearly stable for all Reynolds numbers \citep[page
  132]{drazin1981}.

\subsection{Model derivation}

To derive the model we begin with the 9-mode ansatz of \cite{moehlis1}
but remove the prescribed $z$-dependence. Following \cite{dawes2011}
these modes are rewritten in a mean, toroidal and poloidal form,
$$\mathbf{u}={u_M}(y,z,t)\mathbf{\hat{x}} + \nabla \times
\mathbf{\Phi}_T(x,y,z,t) + \nabla \times \nabla \times
\mathbf{\Phi}_p(x,y,z,t),$$ where these components are defined as
\begin{align*}
\mathbf{u}_m &= \left( A_1 \left( z,t \right) \sin \left( \beta y \right) + A_2 \left( z,t \right) \cos ^2 \left( \beta y \right) \right. \\
&+ \left. A_9 \left( z,t \right) \sin \left( 3 \beta y \right) \right) \mathbf{e_x}, \\
\mathbf{\Phi}_T 	&= A_3 \left( z,t \right) \cos \left( \beta y \right) \mathbf{e_x} + \left( A_4 \left( z,t \right) \sin \left( \alpha x \right) \cos ^2 \left( \beta y \right) \right. \\ 
                  &- A_5 \left( z,t \right) \cos \left( \alpha x \right) \sin \left( \beta y \right) - A_6 \left( z,t \right) \cos \left( \alpha x \right) \cos ^2 \left( \beta y \right) \\ 
			&+ \left. A_7 \left( z,t \right) \sin \left( \alpha x \right) \sin \left( \beta y \right) \right) \mathbf{e_y} ,\\
\mathbf{\Phi}_p &=  A_8 \left( z,t \right) \cos \left( \alpha x \right) \cos \left( \beta y \right)  \mathbf{e_y}.
\end{align*}
leading to
\begin{widetext}
{\footnotesize
\begin{equation}
\begin{array}{lll}
\mathbf{u}_1 = 
\begin{pmatrix}
A_1 \sin ( \beta y ) \\ 
0 \\ 
0
\end{pmatrix}
&
\mathbf{u}_2 = 
\begin{pmatrix}
A_2 \cos ^2 ( \beta y ) \\ 
0 \\ 
0
\end{pmatrix}
&
\mathbf{u}_3 = 
\begin{pmatrix}
0 \\ 
A_3 ' \cos ( \beta y ) \\ 
\beta A_3 \sin ( \beta y )
\end{pmatrix}
\\
&
\\
\mathbf{u}_4 = 
\begin{pmatrix}
 - A_4 ' \sin ( \alpha x ) \cos ^2 ( \beta y ) \\ 
0 \\ 
\alpha A_4 \cos ( \alpha x ) \cos ^2 ( \beta y ) 
\end{pmatrix}
&
\mathbf{u}_5 = 
\begin{pmatrix}
A_5 '  \cos ( \alpha x ) \sin ( \beta y ) \\ 
0 \\ 
\alpha A_5  \sin ( \alpha x ) \sin ( \beta y )
\end{pmatrix}
&
\mathbf{u}_6 = 
\begin{pmatrix}
A_6 '  \cos ( \alpha x ) \cos ^2 ( \beta y )   \\ 
0 \\ 
\alpha A_6  \sin ( \alpha x ) \cos ^2 ( \beta y )  
\end{pmatrix}
\\
&
\\
\mathbf{u}_7 = 
\begin{pmatrix}
-A_7 ' \sin ( \alpha x ) \sin ( \beta y )  \\ 
0 \\ 
 \alpha A_7 \cos ( \alpha x ) \sin ( \beta y )  
\end{pmatrix}
&
\mathbf{u}_8 = 
\begin{pmatrix}
\alpha \beta A_8  \sin ( \alpha x ) \sin ( \beta y ) \\ 
{\mathcal{D}_{\alpha}}^2 A_8 \cos ( \alpha x ) \cos ( \beta y ) \\ 
-\beta A_8' \cos ( \alpha x ) \sin ( \beta y )
\end{pmatrix}
&
\mathbf{u}_9 = 
\begin{pmatrix}
A_9 \sin (3 \beta y ) \\ 
0 \\
0
\end{pmatrix},
\end{array}
\label{modes}
\end{equation}
}
\end{widetext}
where 
$$\mathbf{u} = \sum_{i=1}^9{\mathbf{u}_i}, \enskip A'_i \equiv \frac{{\partial} A_i }{\partial z}, \enskip {\mathcal{D}_{\alpha}}^2 \equiv \left({\alpha}^2 - \frac{{\partial}^2 }{\partial z^2} \right).$$
The Moehlis adaptation introduced $A_9$, adding
variation in $y$ to the base profile, and included $\cos^2( \beta y)$
terms capturing additional nonlinear interactions and linking the new
$A_9$ mode. These modes can be divided into four categories
representing the self-sustaining process introduced by
\cite{waleffe97}. Modes generated by $A_1$ and $A_9$ represent the
mean profile and variation of the mean profile; $A_2$ represents the
downstream streaks which are generated by the roll mode $A_3$; finally
the self-sustaining cycle is closed using five wave-like instability modes
$A_{4-8}$.

Individual evolution equations for the $A_i$ are formed by taking a
Galerkin projection of the Navier-Stokes equations (or its curl) in
$x$ and $y$. The resulting equations have general form
\begin{equation}
\left(\frac{\partial }{\partial t} + \frac{1}{R} L_1 \right) L_2 A_i = N(\mathbf{A},\mathbf{A}) + F,
\label{eqn:general}
\end{equation}
where $L_1$ and $L_2$ are linear operators, $N$ is a quadratic
nonlinear operator and $F$ is the constant forcing term (see appendix
\ref{app:eqn} for details). The nonlinear terms conserve energy as in 
the Navier-Stokes equations.

\subsection{Symmetries}
 
Symmetries play an important part in this work and the system
(\ref{eqn:general}) has several symmetry subspaces. The most important
to the dynamics of this model is inherited from the predecessor models
of Waleffe and Moehlis. This symmetry prescribes the following
$z$-dependence on the modes
\begin{equation}
\mathbf{W} : A_{3\:8}(-z)=-A_{3\:8}(z),
\label{eqn:sym}
\end{equation}
which means $\mathbf{W}:A_3(-z)=-A_3(z)$, $\mathbf{W}:A_8(-z)=-A_8(z)$
with all other $A_i$ unchanged and therefore not shown. Physically,
$\mathbf{W}$ represents the flow field symmetry
\begin{equation}
\left(u,v,w\right)\left(x,y,z\right)=\left(u,v,-w\right)\left(x,y,-z\right),
\end{equation}
for the $x$-independent mean flow modes and 
\begin{equation}
\left(u,v,w\right)\left(x,y,z\right)=\left(-u,-v,w\right)\left(x,y,-z\right),
\end{equation}
for the $x$-dependent instability modes but has no equivalent symmetry
in the Navier-Stokes equations. In this Waleffe system, the symmetry
stems from choosing a roll symmetry, considering the streaks thus
generated and the optimal instability to close the system. Equation
(\ref{eqn:general}) is also invariant under two transformations and
two shifts
\begin{align}
 \mathbf{R}_1 &: (u,v,w)(x,y,z) \rightarrow (-u,-v,w)(\tfrac{1}{2}L_x-x,-y,z),\\
 \mathbf{R}_2 &: (u,v,w)(x,y,z) \rightarrow (-u,-v,w)(L_x-x,-y,z),\\
 \mathbf{R}_1\mathbf{R}_2 &: (u,v,w)(x,y,z) \rightarrow (u,v,w)(x+\tfrac{1}{2}L_x,y,z),\\
 \mathbf{T}_\tau &: (u,v,w)(x,z,z)\rightarrow (u,v,w)(x,y,z + \tau),
\end{align}
which correspond to the following mode operations
\begin{align}
 \mathbf{R}_1 &: A_{2\:3\:4\:5} \rightarrow -A_{2\:3\:4\:5} , \\
 \mathbf{R}_2 &: A_{2\:3\:6\:7\:8} \rightarrow -A_{2\:3\:6\:7\:8} ,  \\
 \mathbf{R}_1\mathbf{R}_2 &: A_{4\:5\:6\:7\:8} \rightarrow -A_{4\:5\:6\:7\:8}, \\
 \mathbf{T}_{\tau} &: \mathbf{A}(z)\rightarrow\mathbf{A}(z + \tau).
\end{align}
 
Spatially, the system is solved using a pseudospectral method based
upon a Fourier series for periodic boundary conditions in sections \ref{sec:modres}
and \ref{sec:local} and Chebyshev polynomials in section \ref{sec:walls} for non-slip sidewalls.
Typical resolutions are 10 Fourier modes per $\pi$ length in the
spanwise direction so that, for example, a steady state in a $10 \pi$
wide domain has 900 degrees of freedom (reduced to 450 by symmetry) and a periodic orbit
three times this.  The equations are time-stepped using Crank-Nicolson
method for the linear terms and a 2nd order Adams-Bashforth method for
the nonlinear terms. Convergence and continuation of simple invariant
sets (e.g. steady states and periodic orbits) was carried out using
PITCON \citep{pitcon}.

\subsection{Spatial Floquet theory}

Starting with a spatially periodic state
\begin{equation}
\mathbf{A}(z)=\sum_{-N}^{N} \mathbf{a}_n e^{i\gamma n z}.
\end{equation}
which has the discrete spatial symmetry $\mathbf{T}_{2\pi/{\gamma}}$ (where $\gamma$ is the solution wavenumber),
disturbances of the form 
\begin{equation}
\mathbf{A}'(z,t)= e^{\lambda t}e^{i\gamma \mu z} \sum_{-N}^{N} \mathbf{b}_n e^{i\gamma n z} + c.c.,
\label{eqn:bif}
\end{equation}
(where $c.c.$ denotes the complex conjugate and thus $\mathbf{A}'$ is
real) were considered in order to uncover Eckhaus (modulational) bifurcations associated with
localization  These disturbances have a wavelength of
$\tfrac{2M\pi}{\gamma}$ which is $M=\tfrac{1}{\mu}$ times longer than
the underlying spatially-periodic state. Only steady bifurcations,
indicated by $\lambda$ passing through 0, were tracked since these were plentiful enough. 
For $M$-odd,
two branches of solution can emanate out of a bifurcation,
corresponding to $\mathbf{A} \pm \epsilon\mathbf{A}' + O(\epsilon^2)$
for $\epsilon \ll 1$ near the bifurcation point. For $M$-even, these
two branches are related by the broken shift symmetry,
\begin{align} \mathbf{T}_{M\pi/{\gamma}}: \mathbf{A}(z) + \epsilon\mathbf{A}'(z) 
&= \mathbf{A}(z + \tfrac{M\pi}{\gamma}) + 
\epsilon\mathbf{A}'(z + \tfrac{M\pi}{\gamma}) \\ \nonumber
&= \mathbf{A}(z) - \epsilon\mathbf{A}'(z).
\end{align} 
There still are 2 branches but they are related by the simple application 
of $\mathbf{T}_{M\pi/{\gamma}}$.


%
%

\section{Steady states and steady modulational instabilities}
\label{sec:modres}

\begin{figure}
\def\svgwidth{\columnwidth}
\SetLabels
(-0.02*0.02){\normalsize $-1$} \\ (-0.01*0.92){\normalsize $1$} \\
(-0.03*0.5){\normalsize $y$} \\ (0.01*-0.1){\normalsize $0$} \\
(0.99*-0.1){\normalsize $2\pi$} \\ (+0.5*-0.15){\normalsize $z$} \\
\endSetLabels \leavevmode
\strut\AffixLabels{\includegraphics[width=0.95\columnwidth,clip=true]{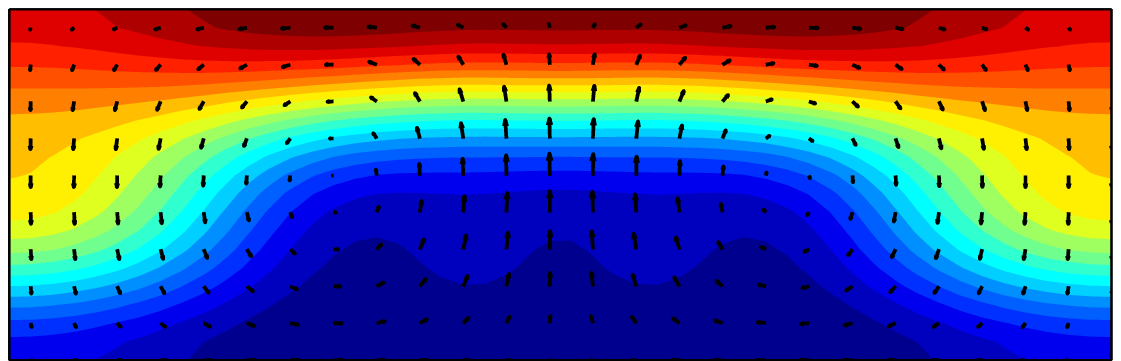}}\\\vspace{5mm}
\def\svgwidth{\columnwidth}
\SetLabels
(-0.02*0.02){\normalsize $-1$} \\ (-0.01*0.92){\normalsize $1$} \\
(-0.03*0.5){\normalsize $y$} \\ (0.01*-0.1){\normalsize $0$} \\
(0.99*-0.1){\normalsize $2\pi$} \\ (+0.5*-0.15){\normalsize $z$} \\
\endSetLabels \leavevmode
\strut\AffixLabels{\includegraphics[width=0.95\columnwidth,clip=true]{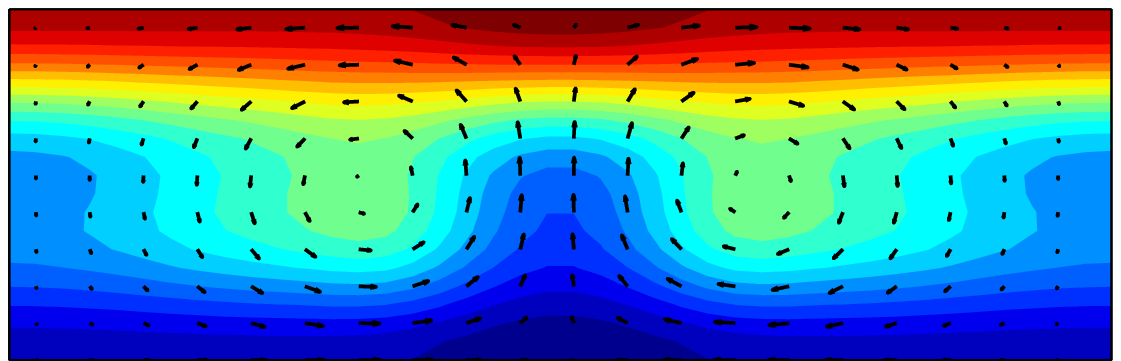}}
\caption{Steady state solutions embedded in `turbulence' at $\Rey=200$ ($S1$: top) and
  $300$ ($S2$: bottom). The plotted flow is $x$-averaged with contours
  denoting flow into (red) and out of the page (blue); and with arrows
  for cross-stream flow. For both solutions the lower branch solutions are plotted.
   The third steady state solution, $S3$, found
  at $\Rey=400$, has a similar appearance (not shown).}
\label{fig:steady}
\end{figure}

To find spatially-periodic solutions from which to study modulational
instabilities, the 9-PDE model was simulated at $\Rey= 200$, $300$ \&
$400$ in a narrow domain $[L_x,L_z]=[4\pi,2\pi]$ where global, long-lived chaotic
behaviour is present.  Velocity fields every 500 time units were
then used as initial conditions in a Newton solver (this `shooting in
the dark' method was only feasible due to the simplicity of the
model). Using this approach, three steady solutions were identified -
hereafter $S1$, $S2$ and $S3$ - all possessing $\mathbf{W}$ symmetry:
see figure \ref{fig:steady}. Each solution was discovered at a
Reynolds number just above their saddle node, with both upper and
lower branches lying in the chaotic saddle. Beyond these Reynolds
numbers, the solutions could no longer be found using this method due to
their increasing instability.
\begin{figure} \SetLabels (+0.0*0.5){\normalsize $E$} \\
(+0.55*-0.02){\normalsize \Rey} \\ \endSetLabels \leavevmode
\strut\AffixLabels{\centerline{\includegraphics[width=0.95\columnwidth]{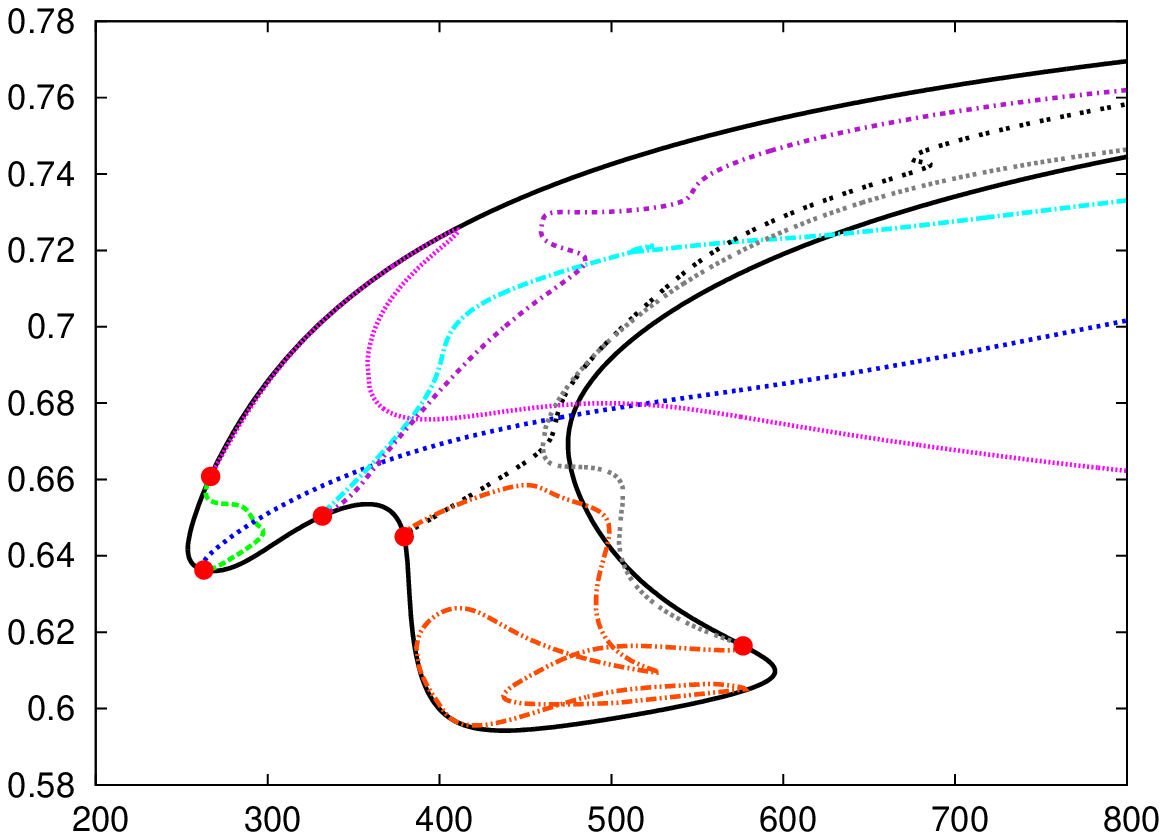}}}
\caption{Continuation curve (black) for the steady state $S2$. Red
dots denote the five $M=3$ modulational bifurcations found for $\Rey \leq
800$ with dashed lines showing the ensuing solution branches which
either rejoin the original branch or extend to large $\Rey$.}
\label{fig:cont3} \end{figure}

Steady modulational bifurcations from these $\mathbf{W}$-symmetric
spatially-periodic states were then searched for. These bifurcations
differ subtly from those in the Swift-Hohenberg equations and the
snaking solutions of \cite{schneider2010}. The periodic solutions in
both these latter settings have two planes of spanwise symmetry
with symmetry about  $z=0$ and antisymmetry about $z=\pi/2\gamma$
which together generate $\mathbf{T}_{2\pi/{\gamma}}$.  Bifurcating
solutions breaking $\mathbf{T}_{2\pi/{\gamma}}$ must then break one of
these reflectional symmetries, generating two solutions, one in each
symmetry subspace.  Here, the symmetry $\mathbf{T}_{2\pi/{\gamma}}$ is
not generated by $\mathbf{W}$ plus a second symmetry plane and
therefore generically modulational bifurcations will remain in the
subspace of $\mathbf{W}$.  Ideally, we should search for bifurcations
from a solution with two symmetry planes, however no such solutions
were found in this model.

\begin{figure} 
\begin{center} 
\SetLabels
(-0.02*0.02){\normalsize $-1$} \\ (-0.01*0.92){\normalsize $1$} \\
(-0.03*0.5){\normalsize $y$} \\ (0.01*-0.07){\normalsize $0$} \\
(0.99*-0.07){\normalsize $6\pi$} \\ (+0.5*-0.1){\normalsize $z$} \\
\endSetLabels \leavevmode
\strut\AffixLabels{\includegraphics[width=0.95\columnwidth,clip=true]{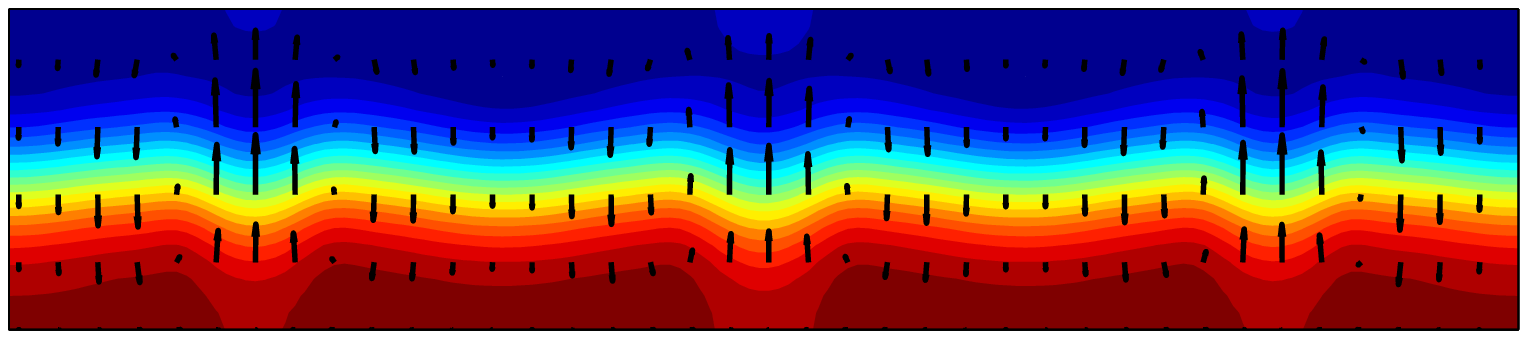}}\\
\vspace{3mm} \SetLabels (-0.02*0.02){\normalsize $-1$} \\
(-0.01*0.92){\normalsize $1$} \\ (-0.03*0.5){\normalsize $y$} \\
(0.01*-0.07){\normalsize $0$} \\ (0.99*-0.07){\normalsize $6\pi$} \\
(+0.5*-0.1){\normalsize $z$} \\ \endSetLabels \leavevmode
\strut\AffixLabels{\includegraphics[width=0.95\columnwidth,clip=true]{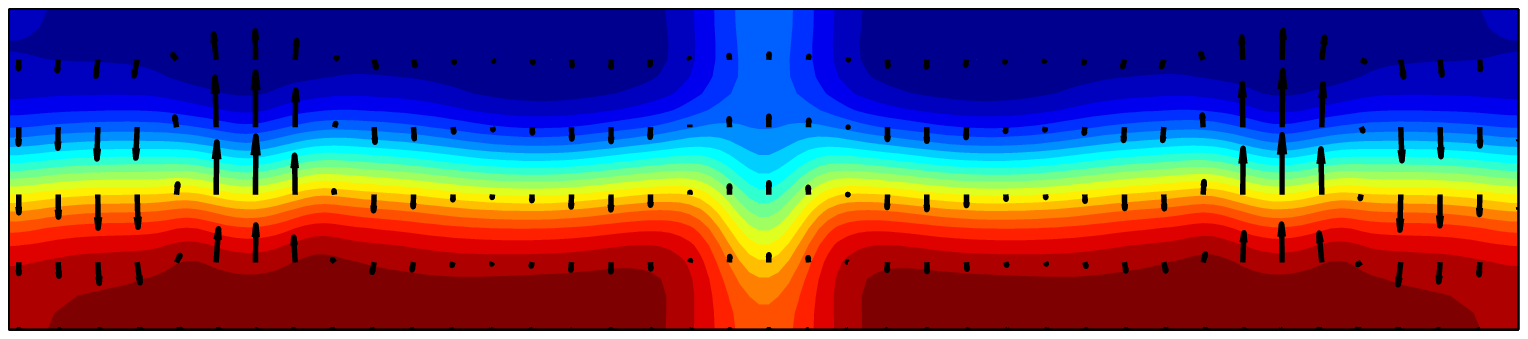}}
\end{center} \vspace{2mm} 

\caption{Steady solutions (velocity deviation from laminar state) at the
bifurcation point from solution $S2$ and at $\Rey=800$ for curve which
emerged through a modulational bifurcation of number $M=3$ (dark blue
dashed line in figure \ref{fig:cont3}). At the bifurcation, the solution is
$T_{L_z/3}$ symmetric, which is broken. At $\Rey=800$ solution remains
global and has similar structure to underlying solution.}
\label{fig:largeRe} \end{figure}

The solution branches emanating from all steady bifurcations with modulation
number $M=3$ from the underlying solution branch, $S2$, are shown in
figure \ref{fig:cont3}: the energy, $E$, defined as 
\begin{equation}
E(t)=\int \tfrac{1}{2} \mathbf{u}^2 dV, 
\end{equation} 
is plotted against \Rey. There are five bifurcations with each
producing two solution branches and their behaviour falls into two
categories: either reconnecting at an alternate bifurcation point, or
continuing to large $\Rey$. In neither case do the solutions localize
let alone snake. A typical solution at large $\Rey$ is plotted in
figure \ref{fig:largeRe} demonstrating the non-trivial flow throughout
the domain (note the aspect ratio of the plots which cause the
transverse velocities to all appear essentially perpendicular to the boundaries).

It is instructive to compare the structure of the bifurcating
eigenfunction with the individual modes in the model at one typical
modulational bifurcation. Figure \ref{fig:eig} plots $A_3$ \& $A_6$,
which represent the rolls and a roll instabilities respectively for
the underlying solution branch (black), and the
modulational (M=3) instability closest to the saddle-node ($A_3'$, $A_6'$ in red). The
eigenfunction $A_3$ is in phase with $A_3$ in the centre of the
domain, and out of phase at the edges of the domain, whereas for $A_6$
the opposite is true. Therefore close to the bifurcation these modes
cannot both act to diminish their respective modes at the same point
in the domain, moving towards a localized solution. This conflicting
behaviour of modes is observed for all of the bifurcations tracked and
helps explain why these bifurcations do not lead to localization.

\begin{figure} 
\includegraphics[width=0.75\columnwidth]{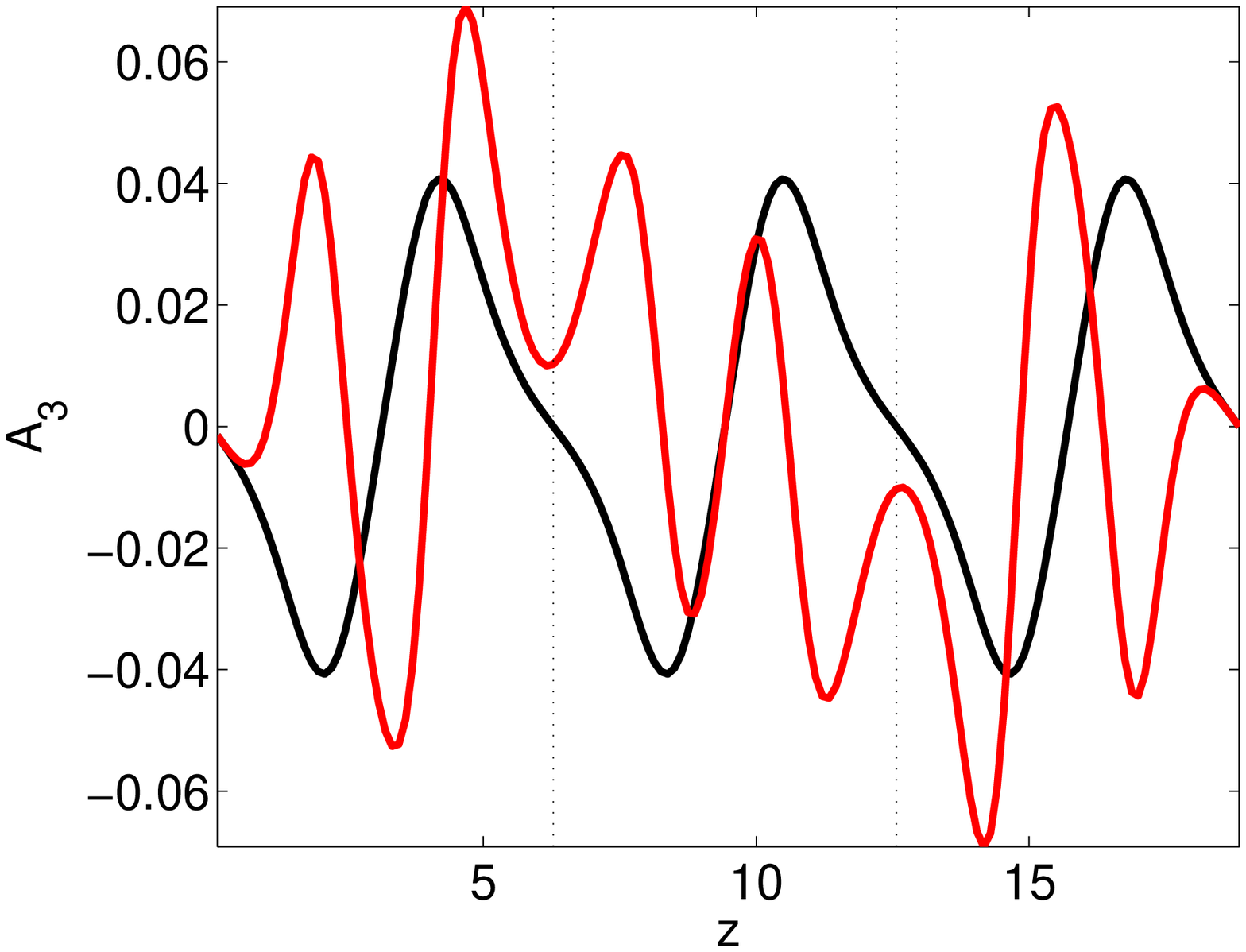}\\
\includegraphics[width=0.75\columnwidth]{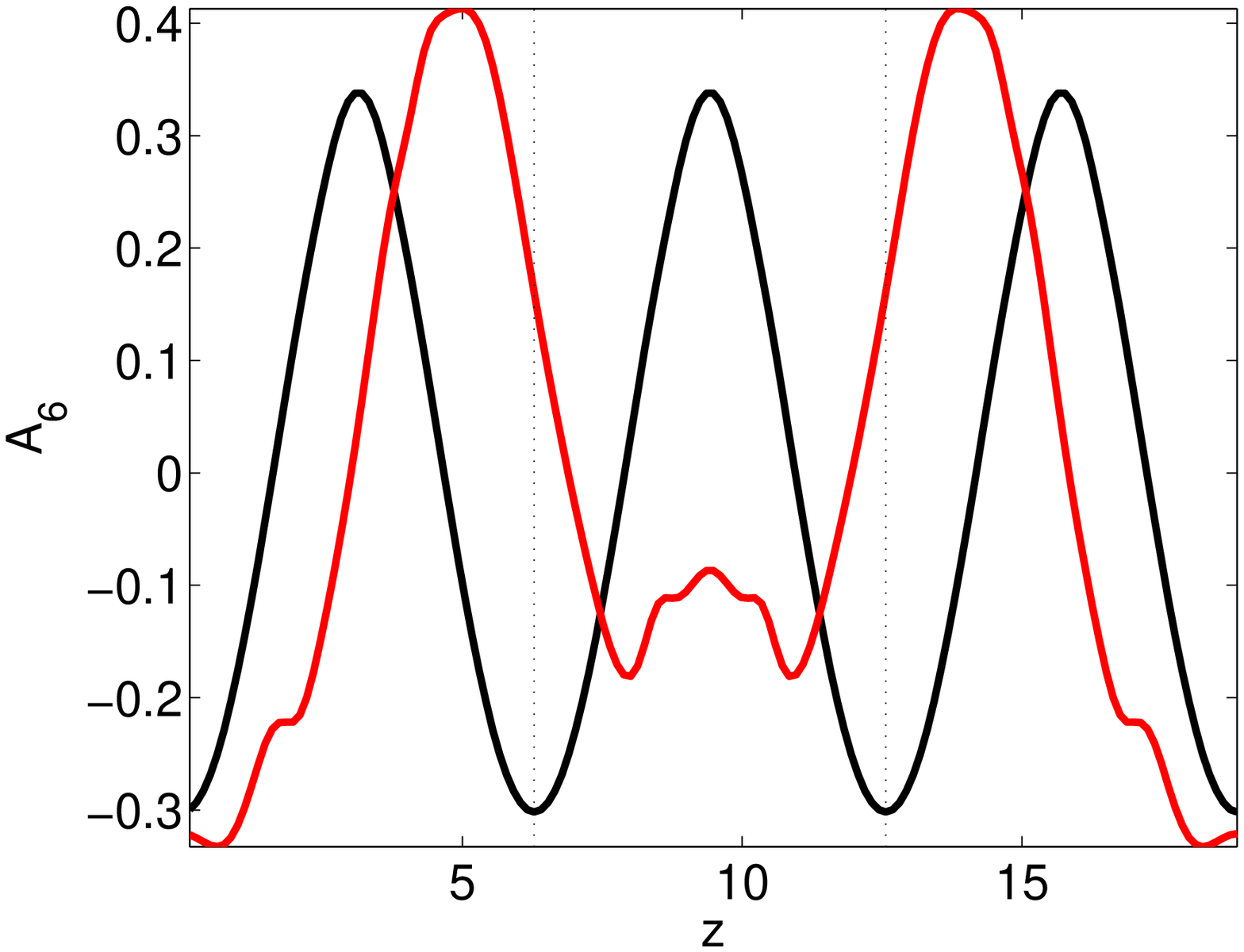} 
\caption{Solution
(black/dark) and eigenfunction (red/light) for two modes ($A_3$ \& $A_6$) at a
$M=3$ modulational instability (blue dashed curve emerging closest to the saddle node). 
The eigenfunction for $A_3$ lies in phase with the underlying solution in the centre of the
domain and out of phase at the edge. For $A_6$, these phases are
switched suggesting that in the linear region of the bifurcation the
eigenfunction cannot move all modes  simultaneously towards localization.}  
\label{fig:eig} 
\end{figure}

Widening the search,  steady modulational instabilities
with $3 \leq M \leq 13$ were sought for $S1$, $S2$ \& $S3$. All
bifurcating solution branches found, however, follow the two
behaviours outlined above, with no evidence of localization. It was
found that bifurcations for $M>5$ were all of a similar type existing
close to one another on the solution curve, as illustrated in figure
\ref{fig:manyBif}. Since no localization was found, the focus was shifted to consider wider domains to
extract localized solutions using a different approach.

\begin{figure}
\SetLabels 
(+0.02*0.5){\normalsize $E$} \\
(+0.55*-0.03){\normalsize \Rey} \\
\endSetLabels 
\leavevmode
\strut\AffixLabels{\centerline{\includegraphics[width=\columnwidth]{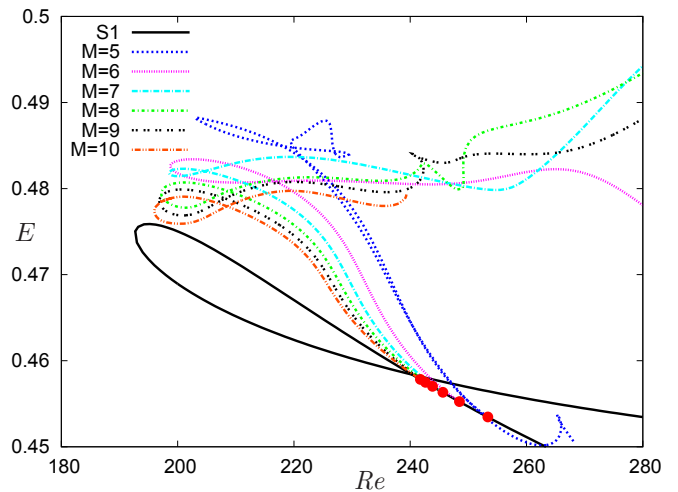}}}

\caption{Continuation curve for solution $S1$ (black curve),
considering the instabilities with similar structure across different
modulation number, $M$. Bifurcations (red dots) lie close together and
initially follow similar behaviour. Bifurcated solutions fail to
localize, irrespective of $M$.}  \label{fig:manyBif} \end{figure}


%
%

\section{Localization in wide domains}
\label{sec:local}

In this section, a different strategy was adopted for generating
localised solutions by considering solutions embedded within the
\textit{edge}. The edge is a hypersurface that separates initial
conditions which become turbulent from those which simply relaminarize
\citep{itano,skufca:edge,schneider:turb} and can be considered a
generalization of the basin boundary for the laminar state to systems
where turbulence is not an attracting state. Despite the unstable
nature of this surface, a bisection technique can be used to track the
dynamics within this surface.  Tracking these dynamics in the various
shear flows has been an effective way to simplify the dynamics and can
lead to exact solutions that are stable within the surface. In pCf,
pipe flow and channel flow, this technique has revealed global
solutions \citep{schneider2008,duguet2008} and the first exact
localized solutions
\citep{schneider2010,avila2013,chantry2013,Zammert2014a,Zammert2014b}.

%
%
\begin{figure} \begin{center} \SetLabels (0.015*0.83){\normalsize $A_1$} \\
(0.015*0.5){\normalsize $A_4$} \\ (0.015*0.17){\normalsize $A_7$} \\
(0.355*0.83){\normalsize $A_2$} \\ (0.355*0.5){\normalsize $A_5$} \\
(0.355*0.17){\normalsize $A_8$} \\ (0.69*0.83){\normalsize $A_3$} \\
(0.69*0.5){\normalsize $A_6$} \\ (0.69*0.17){\normalsize $A_9$} \\ \endSetLabels
\leavevmode
\strut\AffixLabels{\includegraphics[width=0.85\columnwidth]{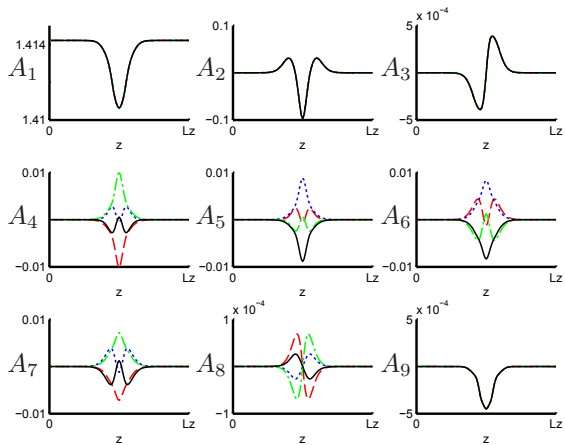}}
\end{center} 
\caption{Localized edge periodic orbit, $P1$. Amplitude of the modes
as a function of $z$. The $x$-independent modes 1-3 and 9 remain steady
(see figure \ref{fig:POLog}), while $x$-dependent instability modes
4-8 have simple fluctuations with a single frequency. Time dependent
modes are plotted at 4 points in the oscillation with time progressing 
through black (full), red (dashed), blue (dotted) and green (dash-dotted).  
} \label{fig:POMod}
\end{figure}

%
%
\begin{figure}
\begin{center}
\SetLabels 
(-0.1*0.55){\normalsize $\smash{\displaystyle\max_{i} a^{(i)}_{n\;m}}$} \\
(+0.55*-0.02){\normalsize $n$} \\
\endSetLabels 
\leavevmode
\strut\AffixLabels{\includegraphics[width=0.75\columnwidth]{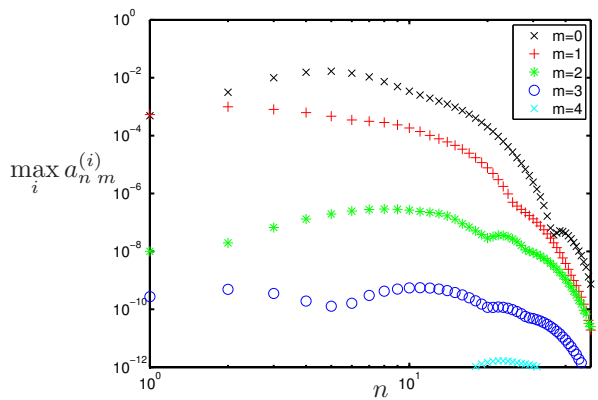}}
\end{center}
\caption{The spectrum (maximum amplitude across all modes) for each
temporal harmonic, where $a^{(i)}_{n\;m}$ is defined in equation
\ref{eqn:timdep}. The temporally constant terms and first harmonic
have similar amplitude whereas the next harmonics have significantly
lower amplitude and can be neglected.  } 
\label{fig:POLog} \end{figure}

Edge-tracking in a wide domain $[L_x, L_z]=[4\pi,10\pi]$ (5 times
wider than in section \ref{sec:modres}) at $\Rey=400$ leads to a
spanwise-localized periodic orbit (hereafter called `P1') depicted in
figure \ref{fig:POMod}. This solution has a short period
($T\approx12$) with a spatiotemporal symmetry 
\begin{align}
\mathbf{u}(x,y,z,t)&=\mathbf{R}_1\mathbf{R}_2\mathbf{u}(x,y,z,t+\tfrac{1}{2}T)\\ \nonumber
&=\mathbf{u}(x+\tfrac{1}{2}L_x,y,z,t+\tfrac{1}{2}T).
\label{eqn:STS} 
\end{align} 
Working with the general representation 
\begin{equation}
A_{j}(z,t)= \sum_{n=-N}^{N} \sum_{m=-MM}^{MM} a^j_{n\;m} e^{i \omega m
t}e^{i\gamma n z},
\label{eqn:timdep} 
\end{equation} 
the structure of the equations is such that the $x$-independent modes of $P1$ 
only possess even powers of $e^{i \omega t}$ and $x$-dependent modes only odd powers, that is,
\begin{align} 
a^j_{n\;2m+1} = 0, \ &j=1,2,3,9,\\ a^j_{n\;2m} = 0, \
&j=4,5,6,7,8.  
\end{align} 
Moreover, the rapid decay of the temporal spectrum - see figure
\ref{fig:POLog} - actually allows for the convergence and accurate
continuation of $P1$ using the extreme temporal truncation
\begin{equation} A_{j}(z,t)= \sum_{n=-N}^{N}
\sum_{m=-1}^{1} a^j_{nm} e^{i \omega m t} e^{i\gamma n z}.
\label{eqn:timapp} 
\end{equation}
$P1$ is also $\mathbf{W}$-symmetric which was not imposed during the edge tracking procedure.

%
%
\begin{figure} 
\SetLabels 
(+0.05*0.85){\normalsize \bfseries{(a)}} \\ (+0.01*0.5){\normalsize $\overline{E}$} \\ (+0.55*-0.035){\normalsize $\Rey$} \\
(+0.9*0.24){\normalsize \bfseries{(b)}} \\ (+0.90*0.72){\normalsize \bfseries{(c)}} \\ \endSetLabels \leavevmode
\strut\AffixLabels{\centerline{\includegraphics[width=0.98\columnwidth]{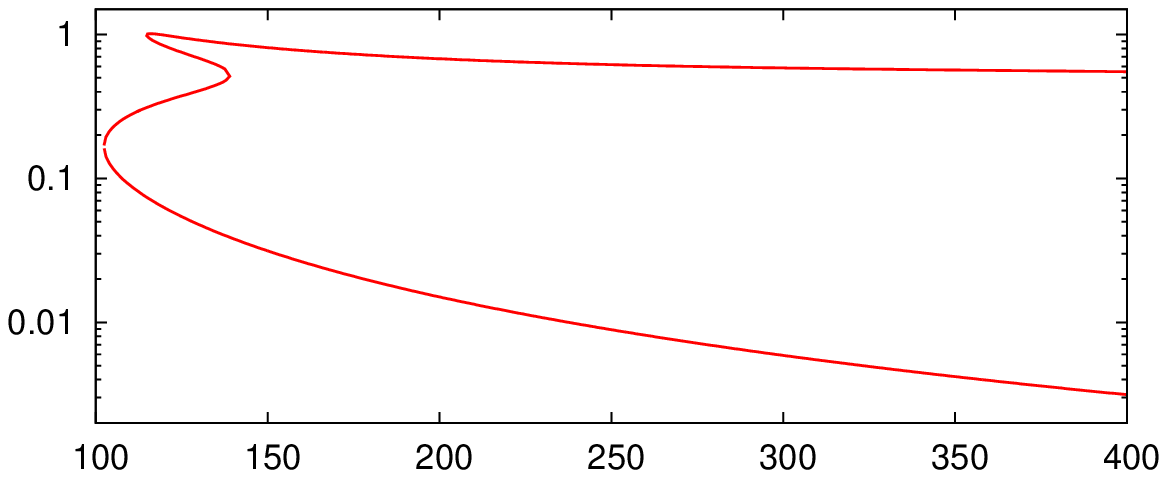}}}
\SetLabels (+0.065*0.67){\normalsize \bfseries{(b)}} \\ (-0.01*+.38){\normalsize
LB} \\ \endSetLabels \leavevmode
\strut\AffixLabels{\centerline{\includegraphics[width=0.95\columnwidth]{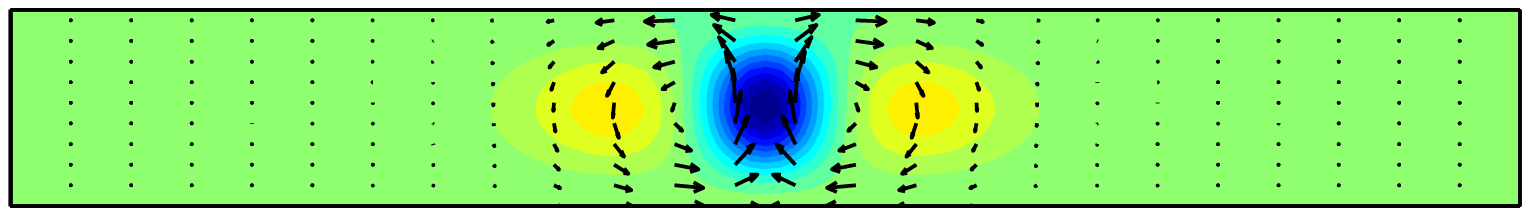}}}
\SetLabels (+0.065*0.67){\normalsize \bfseries{(c)}} \\ (+0.04*-0.25){\normalsize
$0$} \\ (+0.96*-0.25){\normalsize $10\pi$} \\ (-0.00*+.82){\normalsize $1$} \\
(-0.01*+.38){\normalsize UB} \\ (-0.01*-0.06){\normalsize $-1$} \\ \endSetLabels
\leavevmode
\strut\AffixLabels{\centerline{\includegraphics[width=0.95\columnwidth]{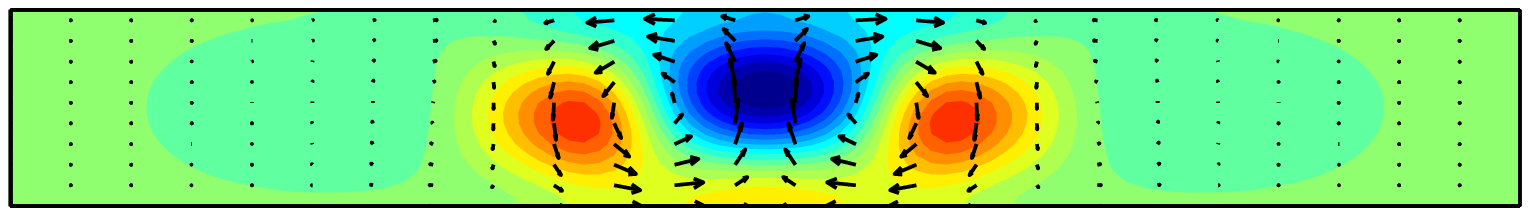}}}\\
\vspace{1mm} 
\caption{Continuation in $\Rey$ of the localized edge
state $P1$ for $L_z=10 \pi$.
Frame (a): the time-averaged energy (see equation (\ref{E_bar})) against
$\Rey$ for the solution, which undergoes three saddle-node
bifurcations before returning to larger Reynolds numbers. The solution
at $\Rey=400$ on the lower and upper branch are shown in frames
(b),(c). During this process the solution increases in width and
amplitude but remains localized.  } \label{fig:cRe} \end{figure}

%
%
\begin{figure}
\SetLabels 
(+0.01*0.9){\normalsize $(a)$} \\
(+0.01*0.5){\normalsize $\overline{E}$} \\
(+0.55*-0.05){\normalsize $L_z$} \\
(+0.88*0.37){\normalsize $\times 2$} \\
\endSetLabels 
\leavevmode
\strut\AffixLabels{\centerline{\includegraphics[width=0.9\columnwidth]{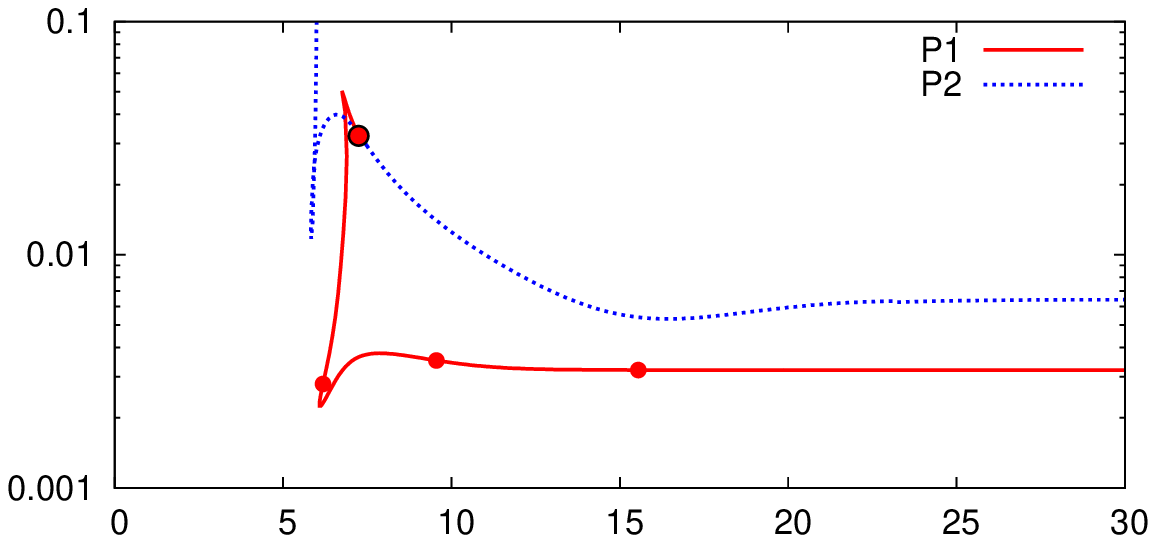}}}\\
\vspace{1mm}
\SetLabels 
(+0.01*0.9){\normalsize $(b)$} \\
(+0.01*0.5){\normalsize $\overline{E}$} \\
(+0.55*-0.035){\normalsize $L_z$} \\
\endSetLabels 
\leavevmode
\strut\AffixLabels{\centerline{\includegraphics[width=0.9\columnwidth]{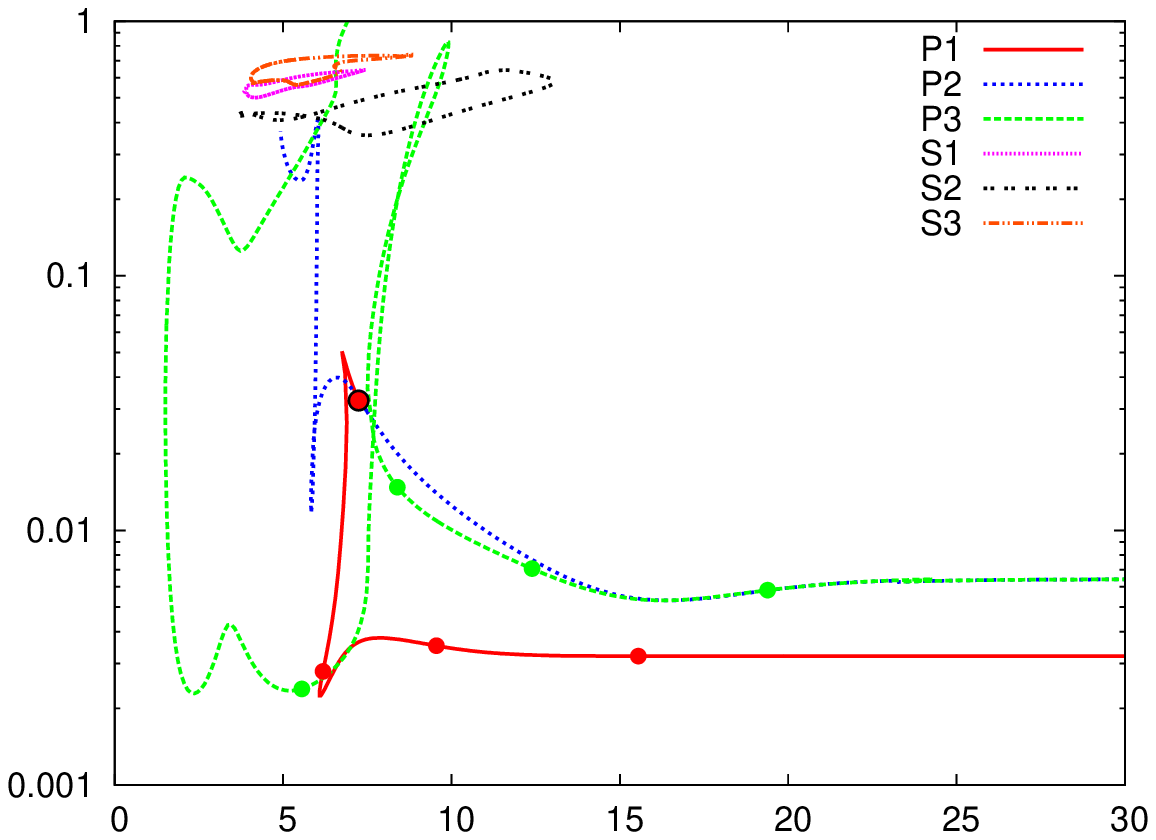}}}\\\vspace{1mm}
\caption{(a) The energy of periodic orbits $P1$ \& $P2$, plotted against
spanwise domain length, $L_z$. The localized single
edge state ($P1$, red) emerges in a symmetry-breaking bifurcation from
solution $P2$ (blue dotted curve, at the black dot). Solution $P2$ can
be continued to wide domains resulting in a symmetric pair of the $P1$
localized state. (b) As above but with the addition of solution $P3$
(green, dashed), the edge state in narrow domains
($L_z<7.5$). Continuation of this solution towards wide domains
results in a localized solution pair with symmetry
$\mathbf{T}_{L_z/2}\mathbf{R}_2$. In contrast, the steady states
$S1$-$S3$ (pink, black and orange) found in the previous section remain global throughout
continuation in $L_z$.}  \label{fig:contPO} \end{figure}

The localized solution of \cite{schneider2010} in plane Couette flow
was found to undergo `snaking' where the localized solution branch
passes through a series of folds with each fold adding a wavelength to
the solution pattern before the domain is filled whereupon a
connection is reached to a spanwise-periodic state. (This behaviour
was originally studied in the Swift-Hohenberg equations
\citep{champneys1998homoclinic,burke2007homoclinic,chapman2009exponential}
and has recently been discovered in other fluid problems
\citep{jaconoporous,beaumeconvection}). To look for snaking behaviour
here, $P1$ was continued in $\Rey$ as shown in figure \ref{fig:cRe}
where the time-averaged energy
\begin{equation}
\overline{E}= \frac{1}{T} \int_0^T E(t) dt, 
\label{E_bar} 
\end{equation} 
is plotted. The continuation curve undergoes 3
saddle-node bifurcations close to the minimum Reynolds number in a
manner similar to snaking curves. However the curve turns back towards
larger values of $\Rey$ and while some structure is added during the
continuation from lower to upper branch the additional structure does
not match the amplitude of the central pattern.  Reintroducing higher
temporal harmonics confirmed these continuation results.

\begin{figure}
\begin{center}\SetLabels 
(-.36*0.5){\normalsize $L_z\approx7$} \\
\endSetLabels 
\leavevmode
\strut\AffixLabels{\includegraphics[height=22mm]{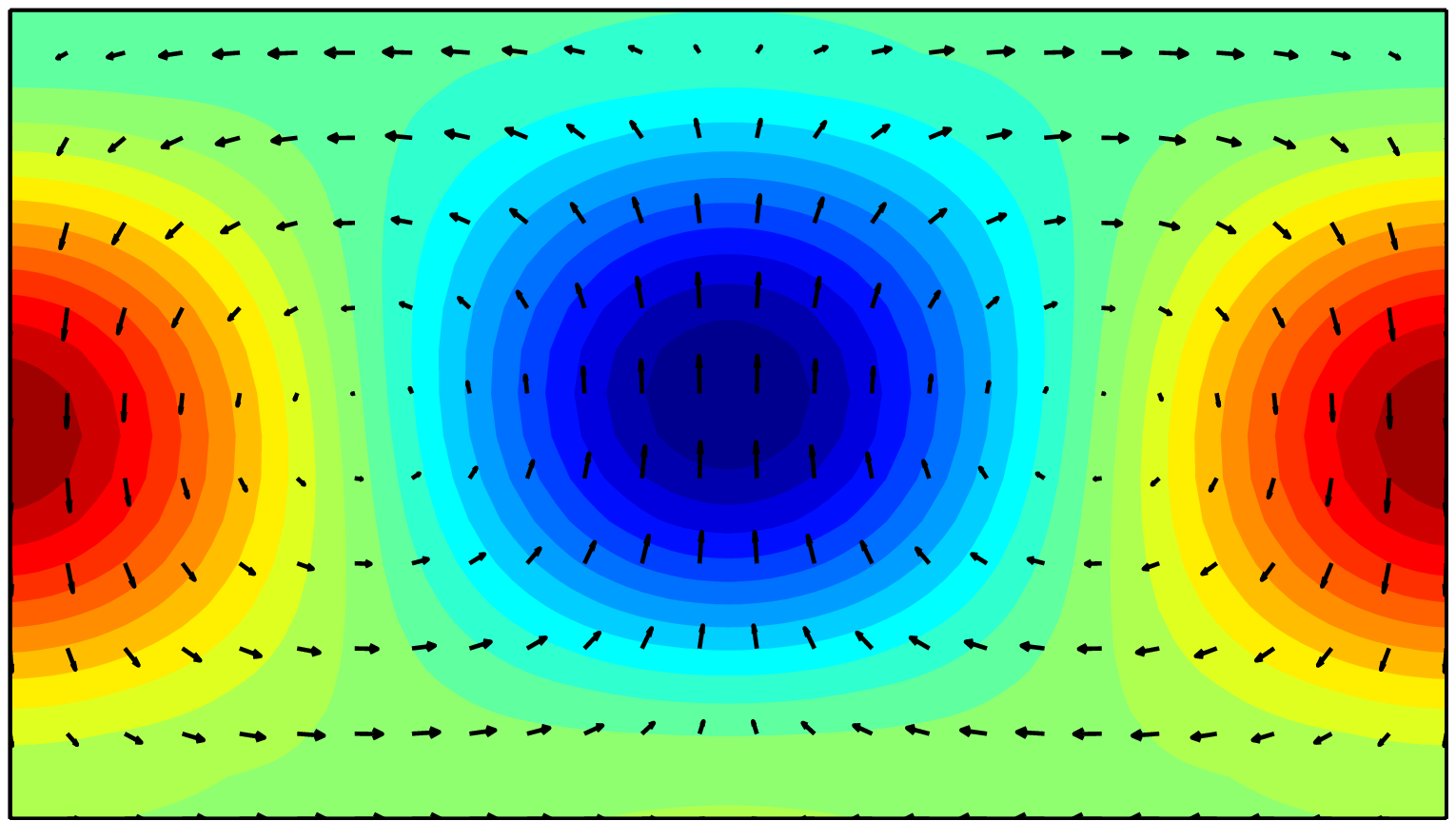}}\\\SetLabels 
(-0.51*0.5){\normalsize $L_z\approx6$} \\
\endSetLabels 
\leavevmode
\strut\AffixLabels{\includegraphics[height=22mm]{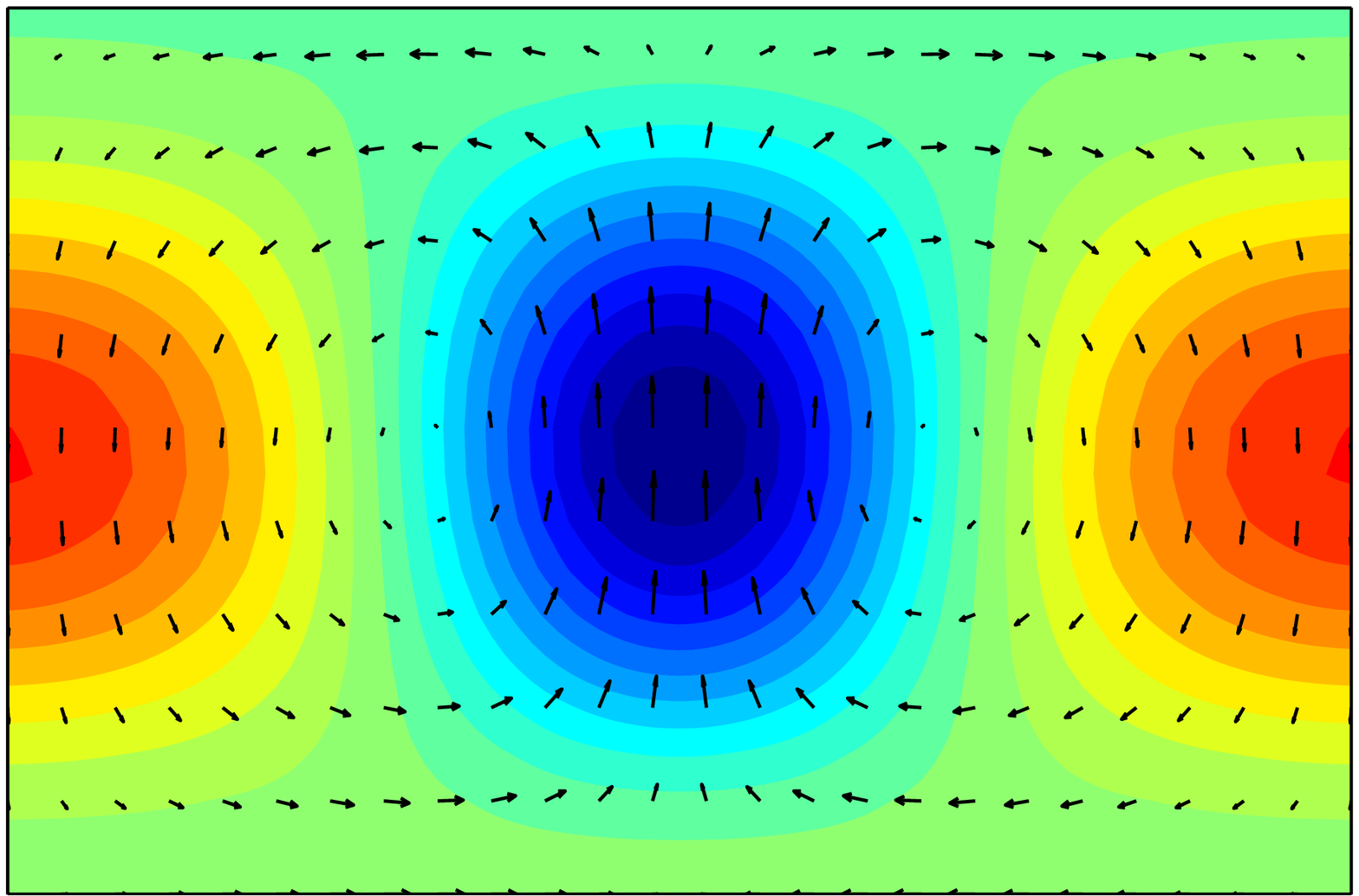}}\\\SetLabels 
(-0.155*0.5){\normalsize $L_z\approx9$} \\
\endSetLabels 
\leavevmode
\strut\AffixLabels{\includegraphics[height=22mm]{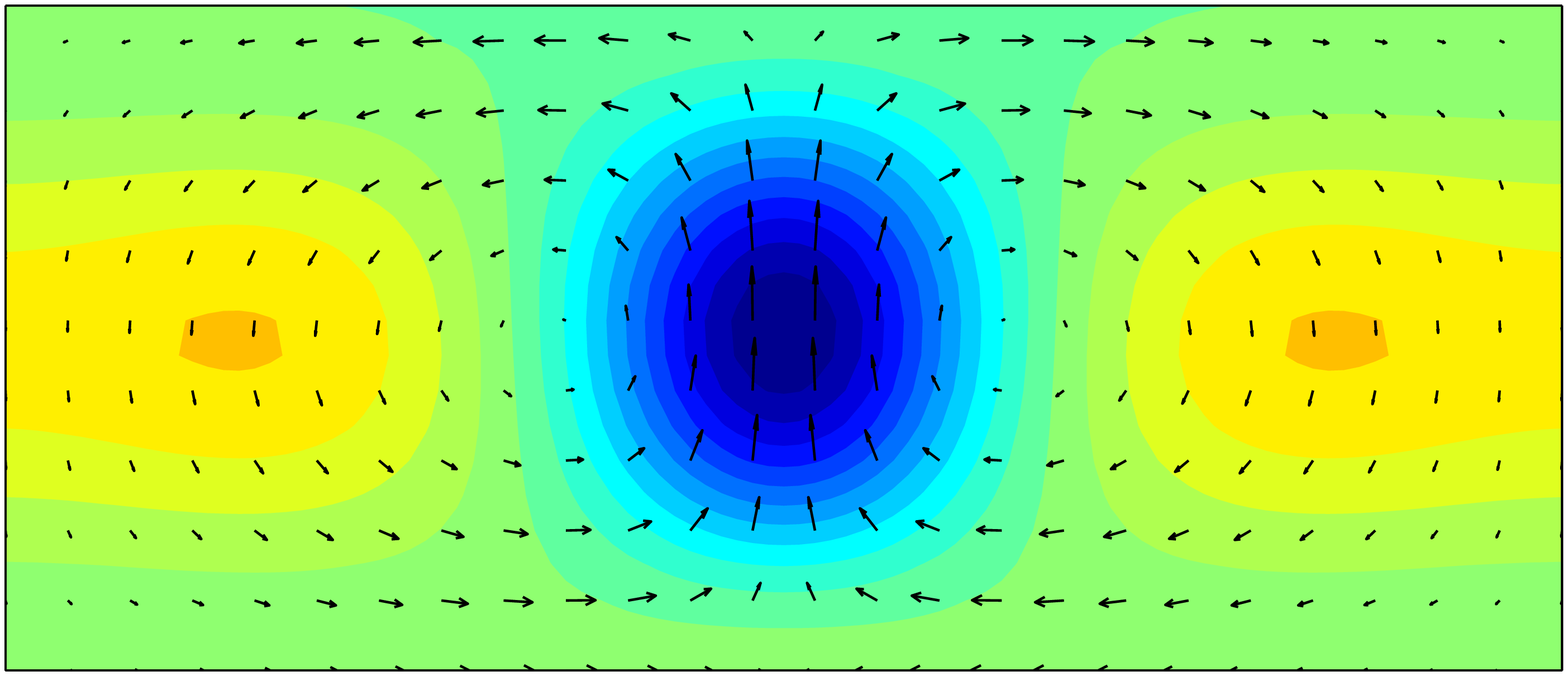}}\\\SetLabels 
(0.1*1.1){\normalsize $L_z\approx15$} \\
\endSetLabels 
\leavevmode
\strut\AffixLabels{\includegraphics[height=22mm]{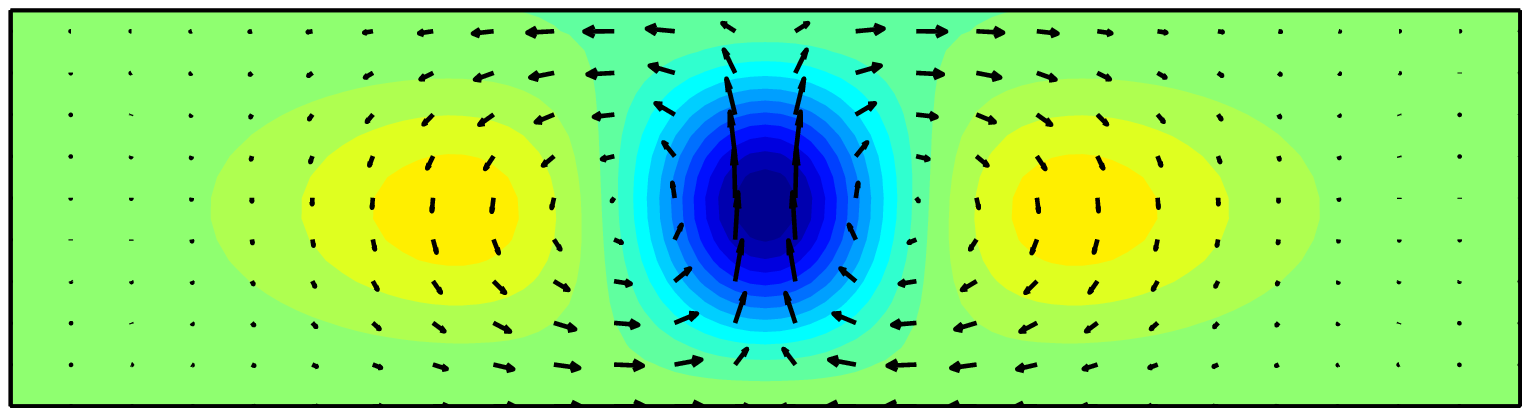}}
\end{center}
\caption{Visualizations of solution $P1$ at varying domain
width corresponding to red dots in figure
(\ref{fig:contPO}). The solutions originates at a symmetry breaking bifurcation 
$\mathbf{T}_{L_z/2}\mathbf{R}_1$ (frame 1), where the streaks are of equal
strength. The slow streak (red) decreases in amplitude 
as domain length decreases. Beyond the saddle-node (in $L_z$) the 
slow streak splits into two and the solution localizes.  Solution 
plotted as $x$-averaged (and thus time-averaged) deviations from 
the laminar state and remains $\mathbf{W}$-symmetric.  } 
\label{fig:local} \end{figure}

Fixing $\Rey$ at 400 and continuing $P1$ to smaller spanwise domain
sizes - see the solid red curve in figure \ref{fig:contPO} -
reveals that $P1$ only starts to feel the spanwise domain size at $L_z
\approx 10$. Below this value, $P1$ undergoes a saddle node
bifurcation at $L_z \approx 6$, after which the time averaged energy
rapidly increases until $P1$ meets a more symmetric state P2 at $L_z
\approx 7.3$ (the solution evolution as a function of $L_z$ is depicted in figure
\ref{fig:local}). $P2$ has the $\mathbf{T}_{L_z/2}\mathbf{R}_1$ symmetry which 
is broken at $L_z \approx 7.3$ to give $P1$.

%
%

\begin{figure}
\begin{center}
\SetLabels 
(+0.01*0.9){\normalsize $(a)$} \\
(+0.01*0.33){\normalsize $(b)$} \\
(+0.02*0.72){\normalsize $\overline{e}(z)$} \\
(+0.02*0.21){\normalsize $\overline{e}(z)$} \\
(+0.55*0.39){\normalsize $z$} \\
(+0.33*+0.01){\normalsize $z$} \\
\endSetLabels 
\leavevmode
\strut\AffixLabels{\includegraphics[width=0.95\columnwidth, clip=true]{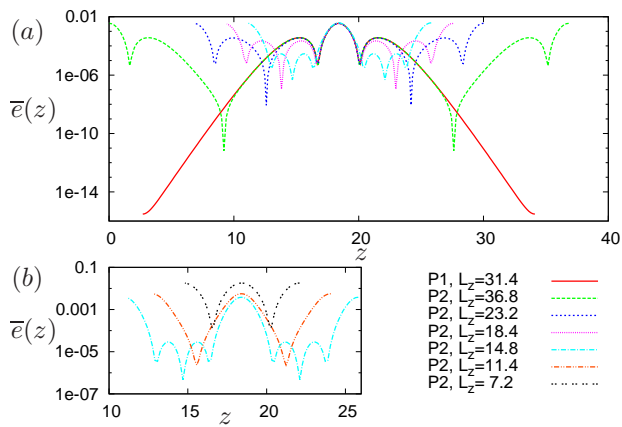}}
\end{center}
\caption{Spanwise energy, $\overline{e}(z)$ (see equation
  \ref{eqn:spweng}) against $z$ for solutions $P1$ (red) and $P2$ (all
  others) as domain width varies. For wide domains (green) $P2$
  matches the exponential decay of localized solution $P1$. During
  continuation to more narrow domains the structure near the energetic
  maxima of $P2$ is unchanged until a domain of $L_z=11.4$ (orange
  dot-dashed). Solutions for $L_z\le 14.8$ are plotted in frame
  (b). At $L_z\approx7.2$ the bifurcation of $P1$ from $P2$ occurs
  (brown dashed).  } \label{fig:LogE} \end{figure}

Continuing this new symmetric solution $P2$ towards larger values of
$L_z$ results in 2 copies of $P1$, related by $\mathbf{R}_1$ and
separated by $\tfrac{1}{2}L_z$. Figure \ref{fig:LogE} shows the
time-averaged energy density as a function of spanwise position
\begin{equation} \overline{e}(z)=\frac{1}{T} \int_0^T \int_0^{L_x}\int_{-1}^{1}\tfrac{1}{2}
\mathbf{u}^2(z,t)\, dxdy, 
\label{eqn:spweng} 
\end{equation} 
For increasing domain widths each localized part of the solution
matches the spatial structure of $P1$ with exponential decay towards
the laminar state at its spanwise extremities ($P2$ has an unstable
eigenvalue of multiplicity two and is therefore unstable within the
edge). $P2$ experiences a saddle node at $L_z \approx 6$ and moving
onto the upper branch the simple temporal ansatz (\ref{eqn:timapp})
begins to break down as higher temporal harmonics become
significant. Figure (\ref{fig:contPO}) seems to indicate that $P2$
  might connect directly to one of the three global states $S1,S2$ or
  $S3$ but the structure of P2 actually precludes this. In all the
  time-periodic solutions found, the $x$-dependent modes are purely
  time dependent, with no time-independent part. Since these modes are
  necessary for non-trivial solutions, the periodic solutions cannot
  be borne directly in a Hopf bifurcation from a steady state.  The
  $x$-dependent modes would first have to develop a time-independent
  component before the time-dependent part could go to zero.  For
  these and computational reasons we did not track the $P2$ solution
  branch further on the upper branch.

%
%

\begin{figure*}
\begin{center}
\SetLabels 
(-2.21*0.5){\normalsize $L_z\approx5.5$} \\
\endSetLabels 
\leavevmode
\strut\AffixLabels{\includegraphics[height=14.75mm]{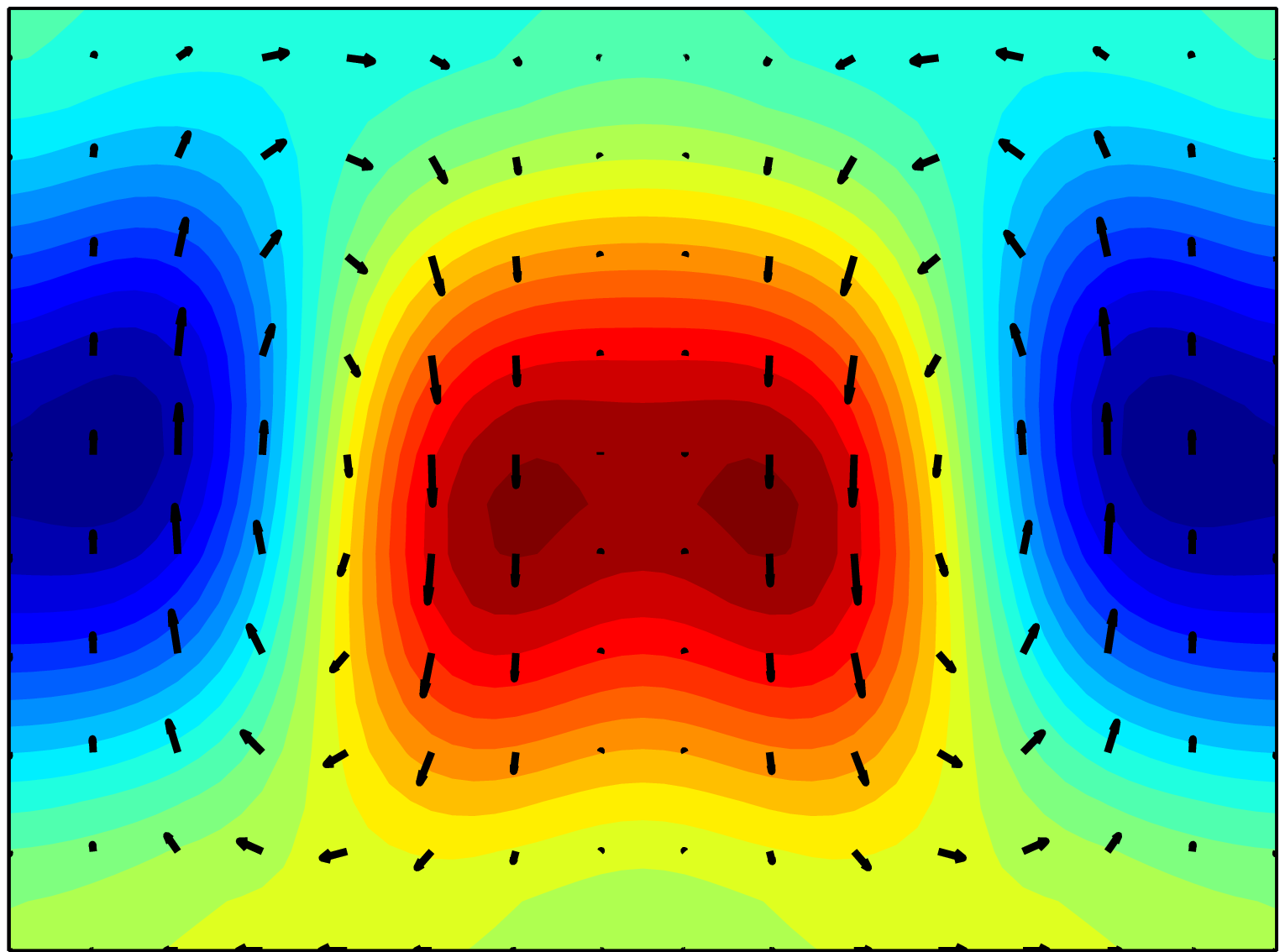}}\\
\SetLabels 
(-1.35*0.5){\normalsize $L_z\approx8.1$} \\
\endSetLabels 
\leavevmode
\strut\AffixLabels{\includegraphics[height=14.75mm]{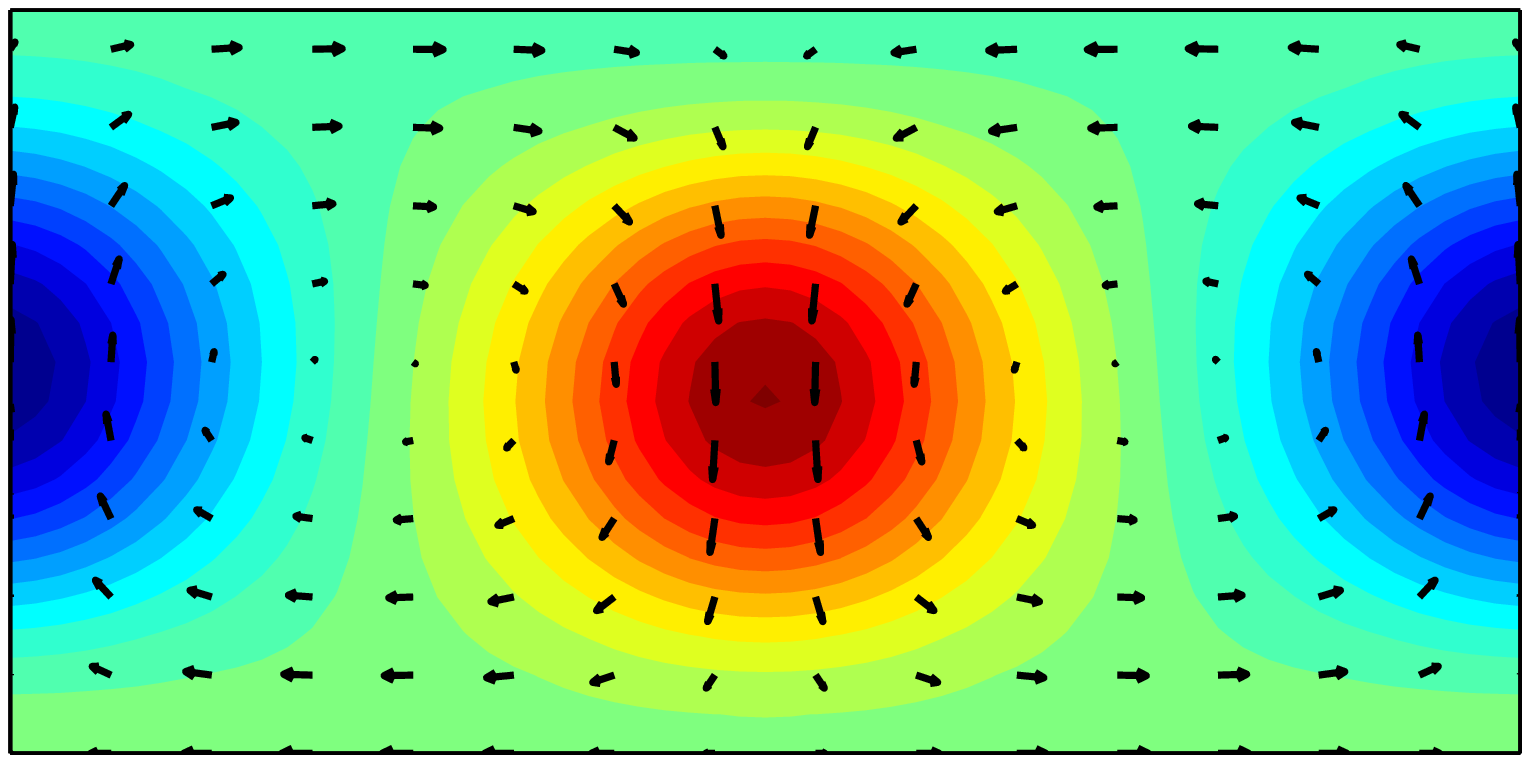}}\\
\SetLabels 
(-0.76*0.5){\normalsize $L_z\approx12$} \\
\endSetLabels 
\leavevmode
\strut\AffixLabels{\includegraphics[height=14.75mm]{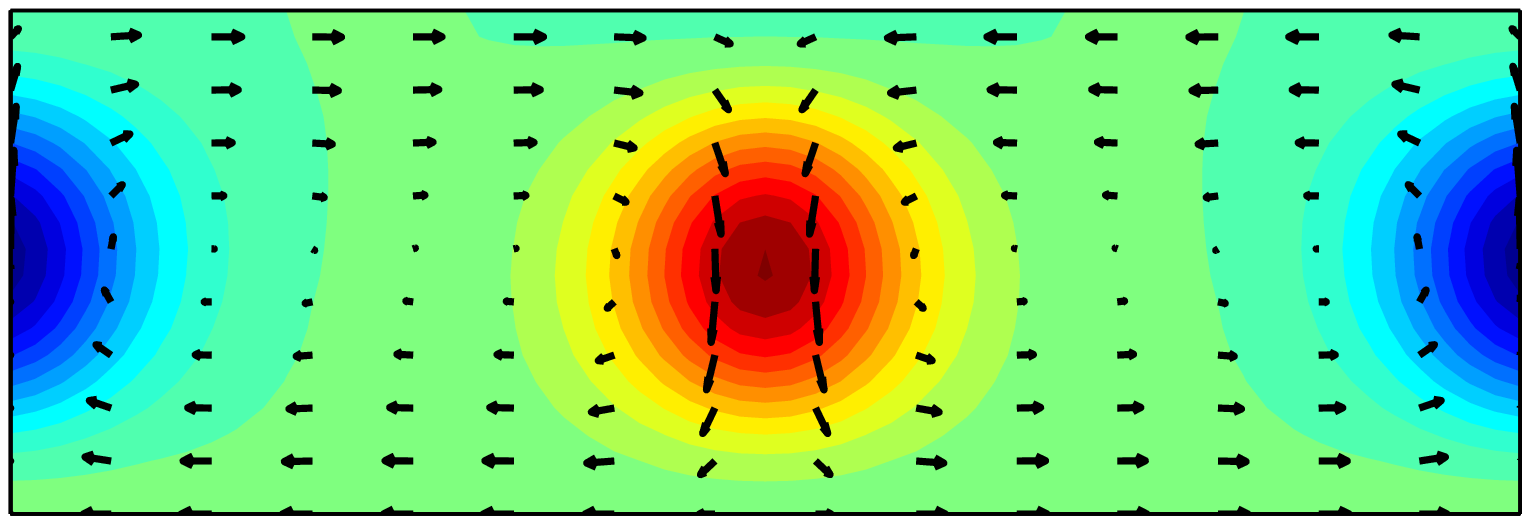}}\\
\SetLabels 
(-0.32*0.5){\normalsize $L_z\approx19$} \\
\endSetLabels 
\leavevmode
\strut\AffixLabels{\includegraphics[height=14.75mm]{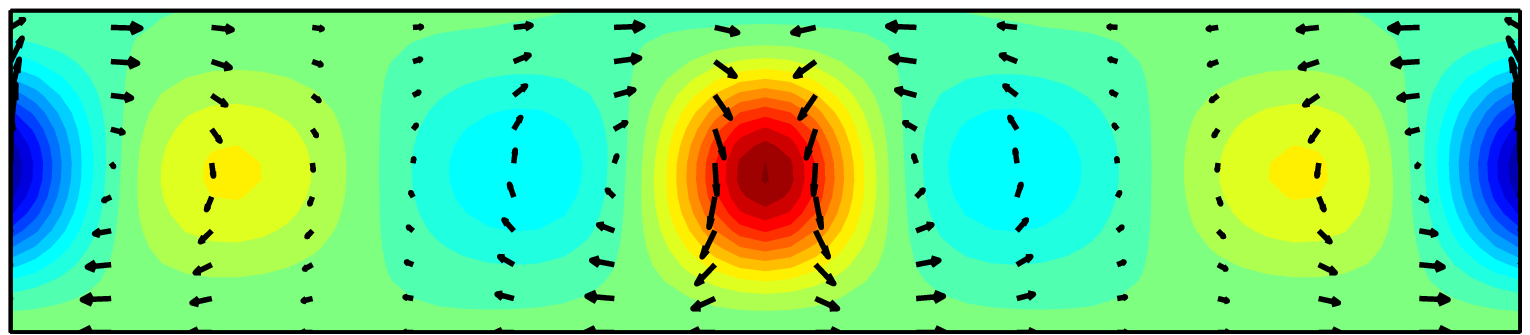}}\\
\SetLabels 
(0.07*1.05){\normalsize $L_z\approx39$} \\
\endSetLabels 
\leavevmode
\strut\AffixLabels{\includegraphics[height=14.75mm]{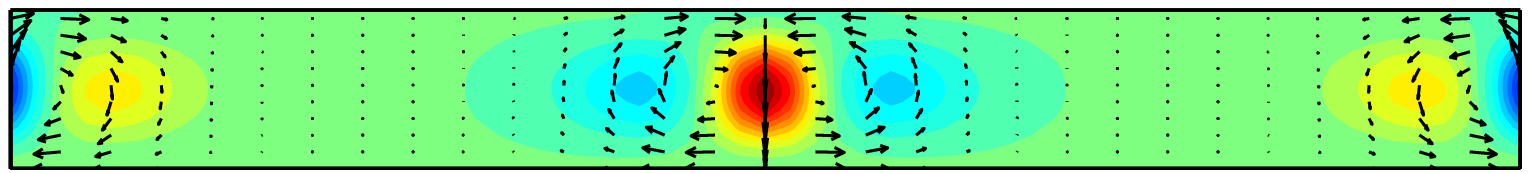}}
\end{center}
\caption{Solution $P3$ during continuation in domain width, $L_z$,
corresponding to green dots in figure \ref{fig:contPO}). Solution is
$\mathbf{T}_{L_z/2}\mathbf{R}_2$-symmetric and forms the attracting
edge state for small domains. As domain width increases ($L_z \approx 7.5$) 
the solution loses stability (and thus is no longer the edge state). 
In wide domains ($L_z > 20$), the solution localizes
as two symmetry related copies of $P1$ seperated by $L_z/2$.  } 
\label{fig:bifI2} \end{figure*}

Edge tracking in smaller domains ($3<L_z<8$) reveals a new
domain-filling periodic orbit, $P3$ as the edge state (see figures
\ref{fig:contPO}(b) and \ref{fig:bifI2}). The solution has simple
temporal structure and two symmetries $\mathbf{W}$ and
$\mathbf{T}_{L_z/2}\mathbf{R}_2$. Continuing this solution to large
domains (dashed green curve in figure \ref{fig:contPO}) the branch briefly
becomes more energetic before stretching out to produce another
localized pair of the periodic orbit $P1$ this time with each $P1$
component being related to the other via the symmetry
$\mathbf{T}_{L_z/2}\mathbf{R}_2$.

%
%

\begin{figure}
\SetLabels 
(-0.09*0.75){\normalsize $P2$} \\
\endSetLabels 
\leavevmode
\strut\AffixLabels{\includegraphics[height=20mm]{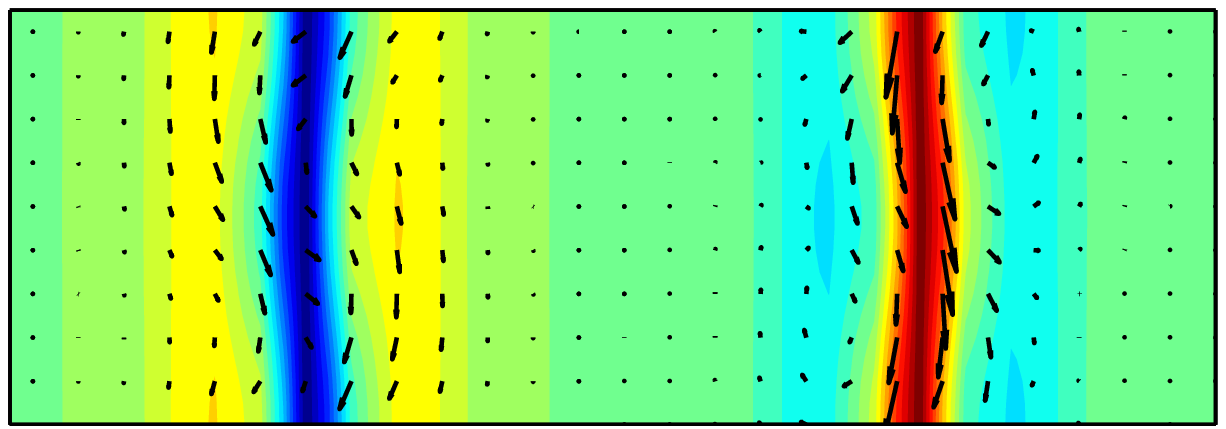}}\\
\SetLabels 
(-0.09*0.75){\normalsize $P3$} \\
\endSetLabels 
\leavevmode
\strut\AffixLabels{\includegraphics[height=20mm]{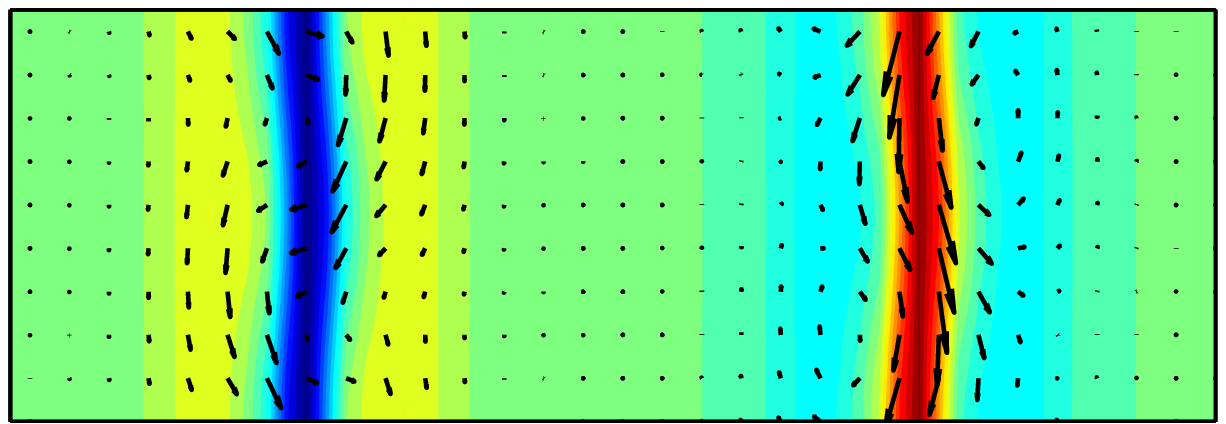}}
\caption{Solutions $P2$ and $P3$ ($L_z \approx 39$) plotted in the
  $(x,z)-$plane at $y=0.9$. Flow in the $y$ direction is plotted with
  contours, while in-plane flow is denoted with arrows. As described
  in equation (\ref{eqn:rel}), solutions are out of phase (both in $x$
  and $t$) in one half of the domain (here the left), while in phase
  in the other half (the right). This subtle difference effects the
  continuation when localized pairs are brought together through
  continuation in $L_z$.  }
\label{fig:XZ} \end{figure}

Localized solution pairs $P2$ ($\mathbf{T}_{L_z/2}\mathbf{R}_1$
symmetric) and $P3$ ($\mathbf{T}_{L_z/2}\mathbf{R}_2$ symmetric)
differ only in the downstream position (or equally, the phase in the
time period,) of one of the individual localized solutions which make
up the pair. The transformation
\begin{equation}
\mathbf{u}(x,y,z,t) \rightarrow
\begin{cases}
\mathbf{u}(x,y,z,t) & \quad z \in \left(0,\tfrac{L_z}{2}\right), \\
\mathbf{u}(x+\tfrac{L_x}{2},y,z,t) & \quad z \in \left(\tfrac{L_z}{2},L_z\right) ,
\end{cases} 
\label{eqn:rel}
\end{equation}
where one localized solution lies in each region, maps one solution
pair to another. This is depicted in figure \ref{fig:XZ} where the two
solutions are plotted in $(x,z)-$plane. This difference in the phase
between the two solutions for large $L_z$ becomes significant as $L_z$ decreases because
the two $P1$ subcomponents  are brought together leading to different global states.

%
%

%
%

\begin{figure} 
\centerline{\includegraphics[width=0.98\columnwidth]{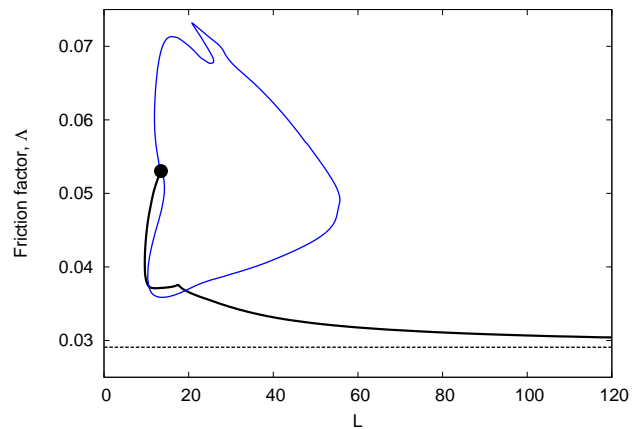}}
\caption{Bifurcation diagram depicting downstream localized pipe flow solution
(black) emerging from a downstream periodic solution (blue) in a
modulational instability. Figure is an adaptation of figure 1 (inset)
from Chantry et al. 2014, here showing friction factor (pressure
gradient down the pipe) against pipe length (in pipe radii).}
\label{fig:chantry} \end{figure}

The behaviour of the solutions $P1$, $P2$ and $P3$ to smoothly move
between an apparently `global' state and a localized structure as the
spanwise domain is changed runs counter to the usual expectation that
there should be a bifurcation between these two extremes. For example,
the localizing states found via edge tracking in \cite{Melnikov2013}
(see their figure 11 (b) and (c)) are the results of a modulational
instability from a global travelling wave state (see their figure 6
and 11(a)). One explanation is that a localised solution
appears to become global near to where it connects to a global
branch of solutions as in \cite{chantry2013}. Figure \ref{fig:chantry}
replots the inset of figure 1 in \cite{chantry2013} to show the
localised (black) branch going through a turning point (corresponding
to the minimum domain size) before connecting to the closed
(blue) global branch in a slightly larger domain (localization here 
is in the streamwise rather than spanwise direction). At this turning point,
the flow on the localised branch looks domain filling. However, the
localised states in our model are not connected to any global state at
the Reynolds numbers considered and there are structural reasons to
expect there not to be such a connection at all (the $x$-dependent
modes have no steady part).  The fact that this behaviour has also
been seen in the preliminary work of \citet[][page 89, fig 5.1]{okino}
on rectangular duct flow indicates that this observation is not an
artefact of the model. In \cite{okino}, a (global) travelling wave solution
found in a square duct was smoothly continued to wide aspect-ratio
ducts to reveal a spanwise-localised travelling wave with very little
adjustment in the structure from the square duct situation.  One
difference between the work here and Okino's calculation is the
spanwise boundary conditions imposed: periodic here and no-slip there.
It is worth confirming that this is not significant.

%
%

\section{Spanwise walls}
\label{sec:walls}
\begin{figure}
\begin{center}
\SetLabels 
(+0.02*0.5){\normalsize $E$} \\
(+0.55*-0.01){\normalsize $t$} \\
\endSetLabels 
\leavevmode
\strut\AffixLabels{\includegraphics[width=\columnwidth, clip=true]{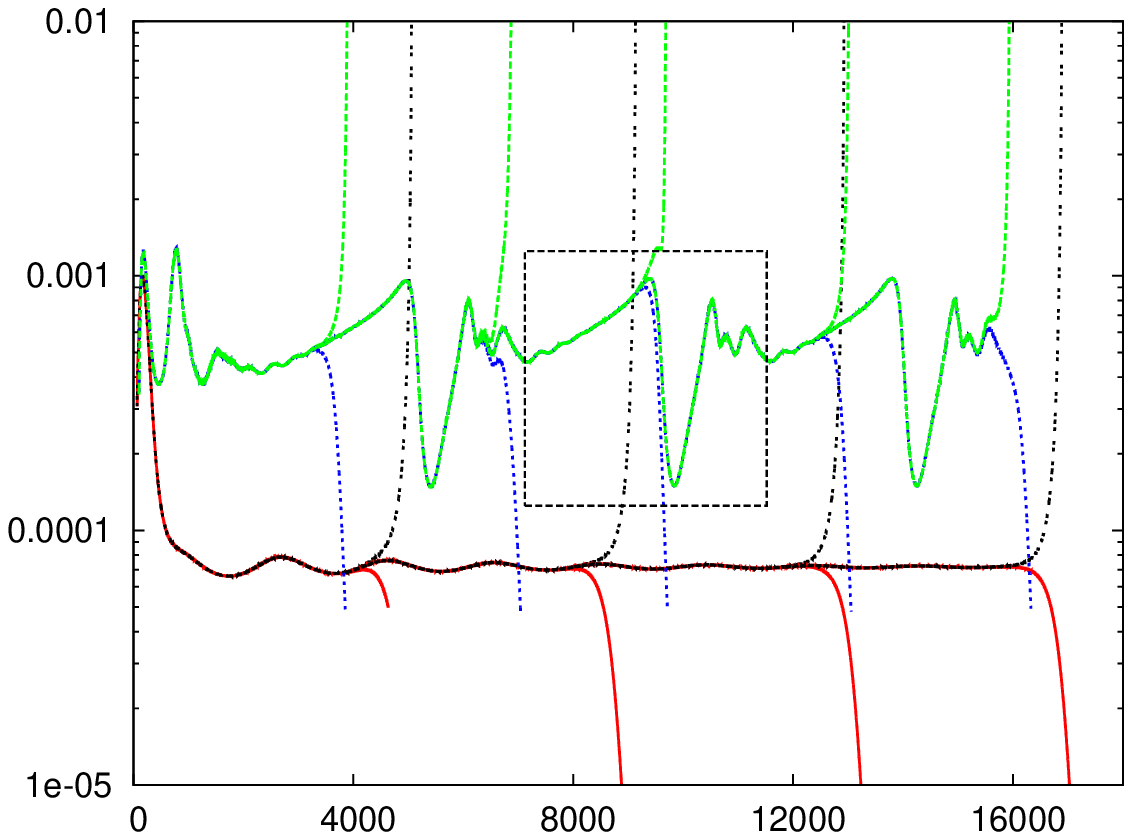}}
\end{center}
\caption{Dynamics within the edge for two global initial conditions
  generated from a turbulent trajectory. The first localizes away from
  the wall and converges towards the low energy localized periodic
  orbit found with periodic boundary conditions. The second condition
  localizes adjacent to the wall to a more complex (both in time and
  space) and higher energy solution. The black box encloses a single
  (half) period which is further studied in figure \ref{fig:WallPO}.
}
\label{fig:WallEdge}
\end{figure}

The periodic spanwise boundary conditions are replaced in this section by no-slip conditions on
the deviations from the laminar flow,
$$\mathbf{u}(x,y,0,t)=\mathbf{u}(x,y,L_z,t)=0,$$
which means the modes satisfy the following conditions
\begin{align}
A_i(0)=A_i(L_z)=0,\enskip &i=1,9,\\
\left.\frac{d A_{i}}{d z}\right|_{z=0}=\left.\frac{d A_{i}}{d z}\right|_{z=L_z}=0,\enskip &i=3,8,\\
\left.\frac{d^2 A_i}{d z^2}\right|_{z=0}=\left.\frac{d^2 A_i}{d z^2}\right|_{z=L_z}=0,\enskip &i=8.
\end{align}
The equations are solved using a Chebyshev collocation
method. Searching for localization, a domain of size
$[L_x,L_z]=[4\pi,16\pi]$ is considered at Reynolds number 400. In a domain of these
dimensions the localized periodic orbit, $P1$, (see frame 1 of figure
\ref{fig:local}) is an edge attractor both for periodic and no-slip
boundary conditions. However, while evidence suggests that the
periodic orbit is the unique attractor for periodic boundary
conditions, this is not true for no-slip conditions. In this domain,
in addition to $P1$ a second solution, $P4$, attracts the dynamics
within the edge. The energetic evolution of two initial conditions in
the edge is plotted in figure \ref{fig:WallEdge}, where the two
initial conditions were generated from turbulent initial conditions
separated by 30 time units. This new solution, $P4$, attracts the
dynamics when the fluctuations are strong near one of the wall regions
and the dynamics along the orbit are shown in figure
\ref{fig:WallPO}. Evidence from edge tracking suggests that the
solution is a periodic orbit where the state after half a period
($\tfrac{T}{2}=4418$) is related to the initial state through
spatial-temporal symmetry
\begin{equation}
\mathbf{u}(x,y,z,t+\tfrac{T}{2})=\mathbf{R}_1\mathbf{u}(x,y,z,t).
\end{equation}

However, the extremely long period of the structure prevents
convergence of the solution. Within the slow time-scale behaviour the
$x$-dependent modes oscillate on a time-scale of 24 units throughout
the period (not visible in figure \ref{fig:WallPO}). Ignoring the
long-timescale dynamics these oscillations take a similar form to
those exhibited by the temporally-simple localized periodic orbit with
$$\mathbf{u}(x,y,z,t+12)\approx \mathbf{R}_1
\mathbf{R}_2\mathbf{u}(x,y,z,t).$$ The transition from quiescent
evolution (the first seven snapshots in figure \ref{fig:WallPO}) to fast evolution (the last five snapshots), 
suggest the solution may be an analog of the bursting solutions of \cite{kawahara,khapko2013}.

\begin{figure}
\begin{center}
\SetLabels 
(+0.02*0.5){\normalsize $E$} \\
(+0.55*-0.03){\normalsize $t$} \\
\endSetLabels 
\leavevmode
\strut\AffixLabels{\includegraphics[width   =\columnwidth, clip=true]{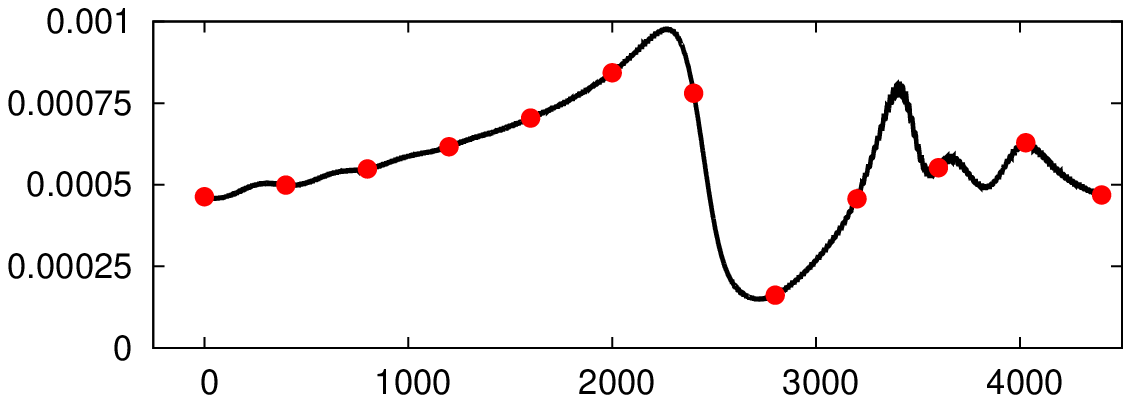}}
\includegraphics[width=0.95\columnwidth, clip=true]{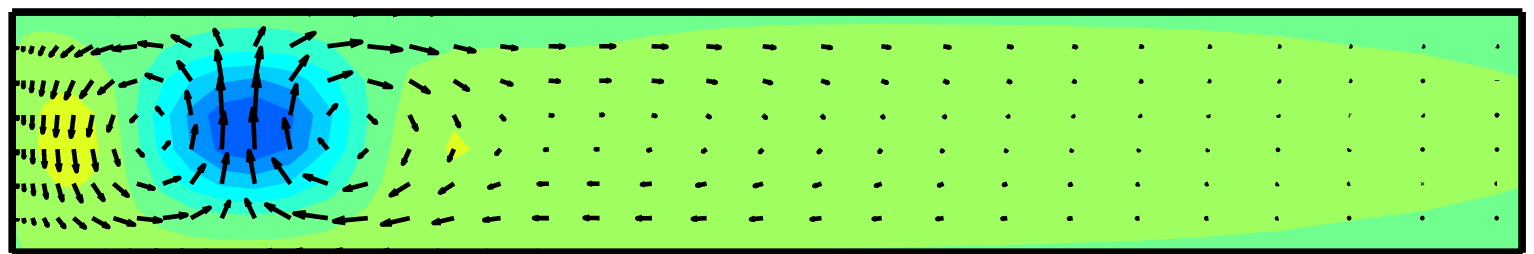}
\includegraphics[width=0.95\columnwidth, clip=true]{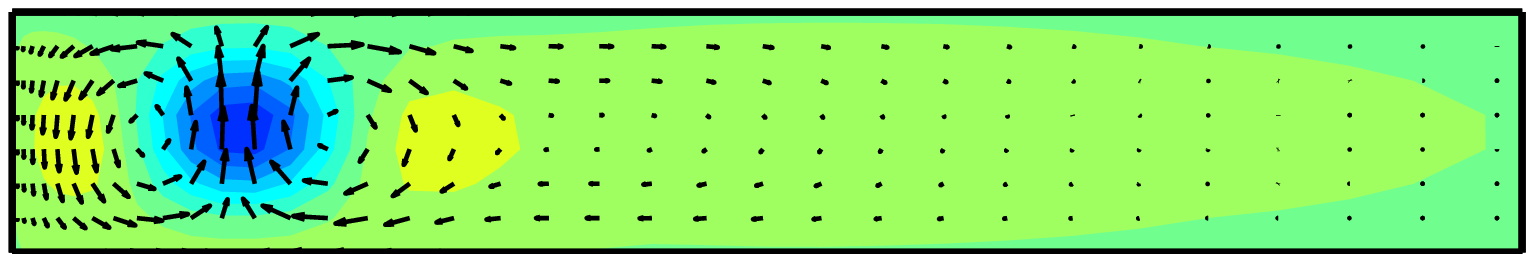}
\includegraphics[width=0.95\columnwidth, clip=true]{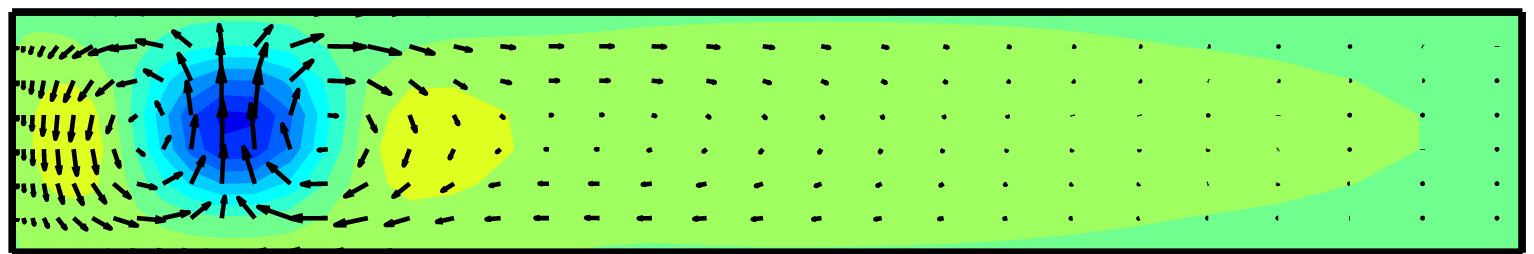}
\includegraphics[width=0.95\columnwidth, clip=true]{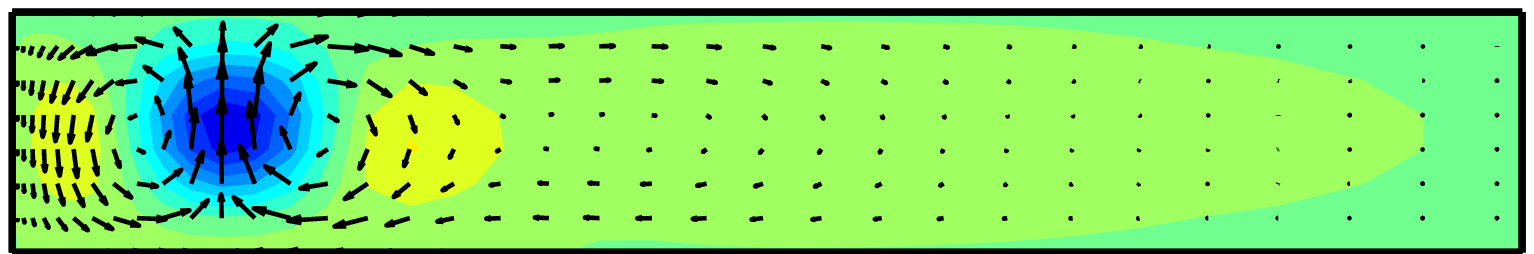}
\includegraphics[width=0.95\columnwidth, clip=true]{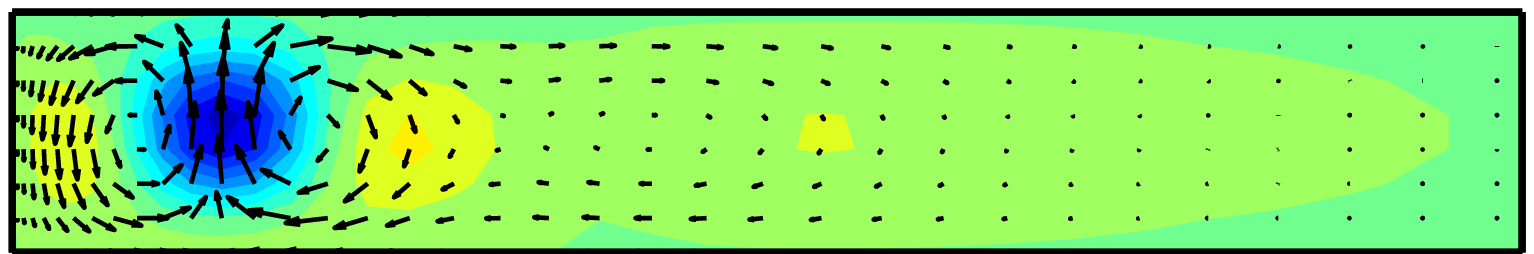}
\includegraphics[width=0.95\columnwidth, clip=true]{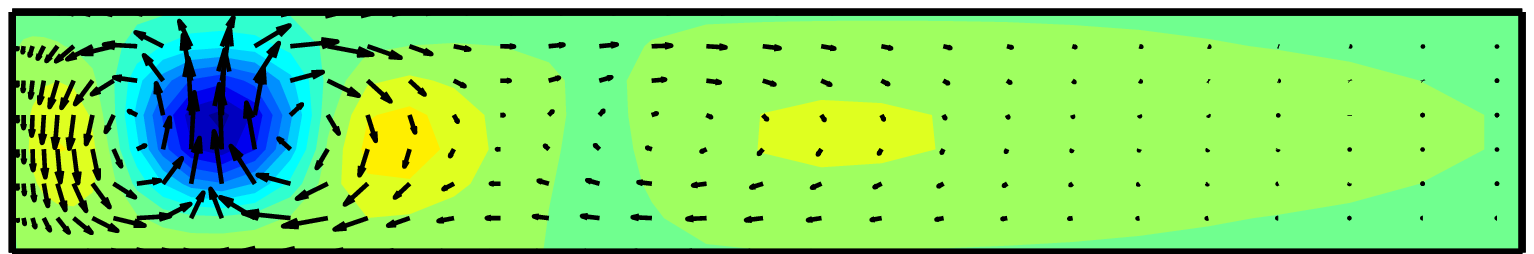}
\includegraphics[width=0.95\columnwidth, clip=true]{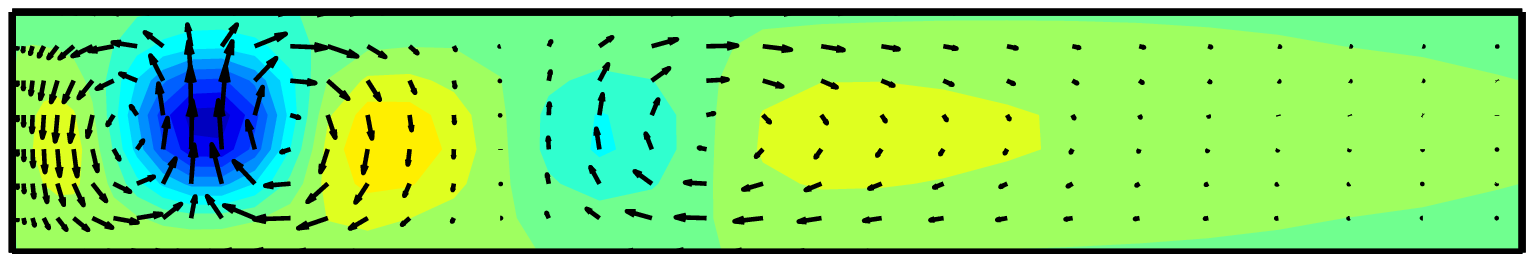}
\includegraphics[width=0.95\columnwidth, clip=true]{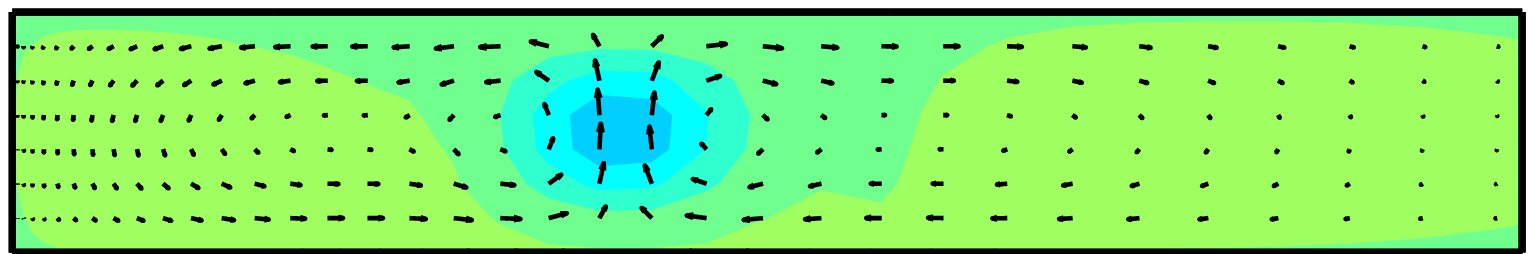}
\includegraphics[width=0.95\columnwidth, clip=true]{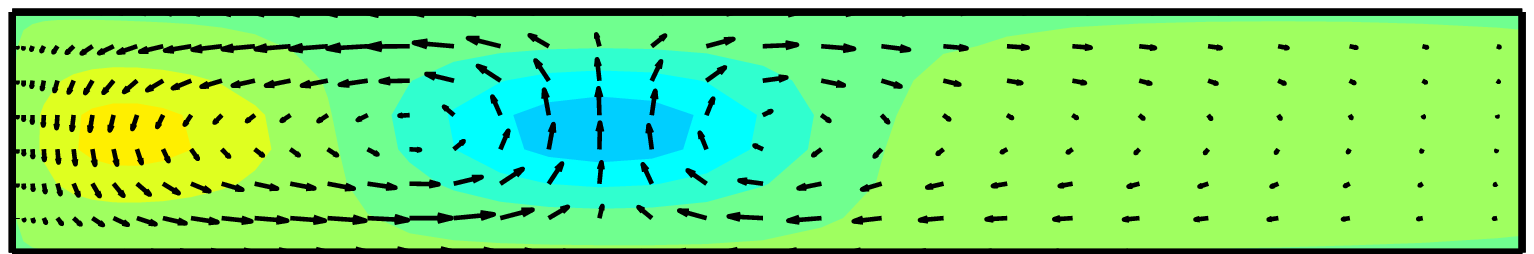}
\includegraphics[width=0.95\columnwidth, clip=true]{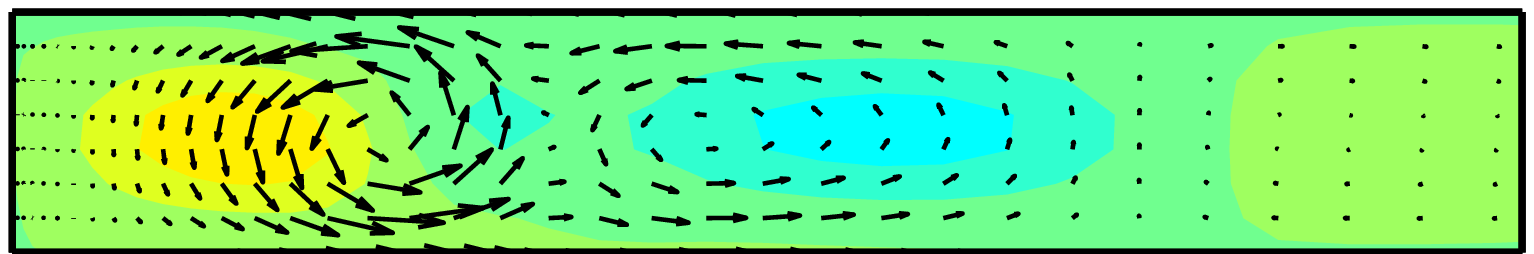}
\includegraphics[width=0.95\columnwidth, clip=true]{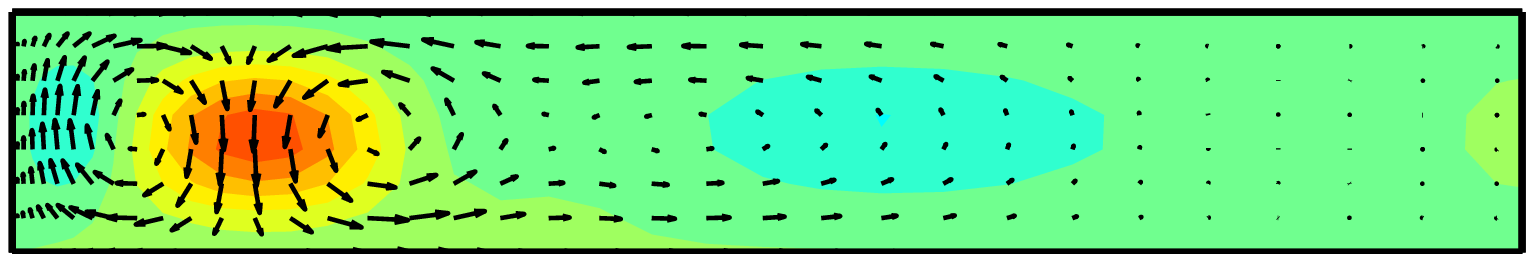}
\SetLabels 
(+0.03*-0.28){\normalsize $0$} \\
(+0.96*-0.28){\normalsize $4\pi$} \\
(-0.01*+.78){\normalsize $1$} \\
(-0.025*-0.10){\normalsize $-1$} \\
\endSetLabels 
\leavevmode
\strut\AffixLabels{\includegraphics[width=0.95\columnwidth, clip=true]{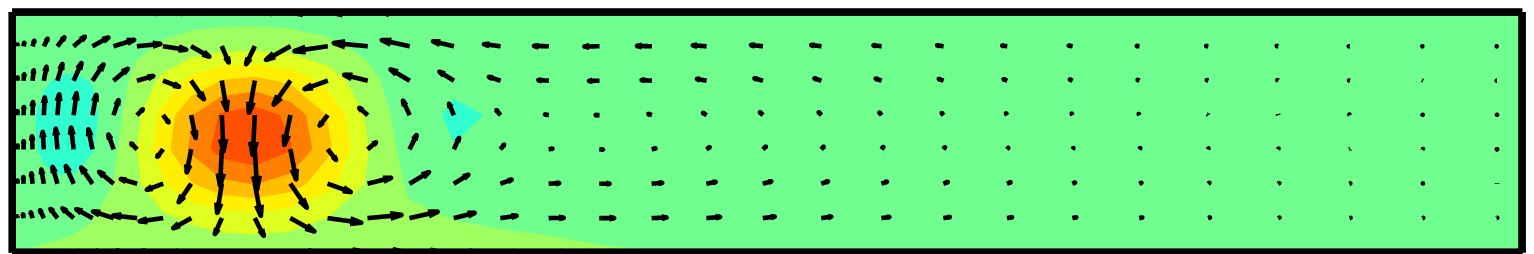}}
\end{center}
\caption{Evolution along an edge attracting periodic orbit, $P4$. Top
  frame shows the energetic evolution along half the period at which
  point the state is related by symmetry $\mathbf{R}_1$ to the initial
  state. Dots correspond to the twelve $x$-averaged flowfield
  snapshots (deviation from laminar flow). No-slip conditions are
  applied at $z=0,16\pi$ with only the left quarter of the domain
  depicted. Solution initially grows slowly in time while moving
  slightly towards the wall before breaking down and moving away from
  the wall. Finally the structure reforms with inverted streak-roll
  structure.  }
\label{fig:WallPO}
\end{figure}

Finally, we consider the continuation of solution $P1$ with
spanwise-walls imposed (figure \ref{fig:P1Wall}). For domains of width
$L_z>10$ the solution is indistinguishable from the periodic boundary
condition equivalent. Below this value the fast streaks (yellow) are
squashed by the incoming walls, while the central slow streak (blue)
remains unaltered so that the smooth transition between localized and domain
 filling solution is again recovered. Beyond the saddle-node, the time complexity
increases along the upper branch and the solutions were not tracked. 

\begin{figure}
\begin{center}
\SetLabels 
(+0.01*0.5){\normalsize $\overline{E}$} \\
(+0.55*-0.01){\normalsize $L_z$} \\
\endSetLabels 
\leavevmode
\strut\AffixLabels{{\includegraphics[width=\columnwidth]{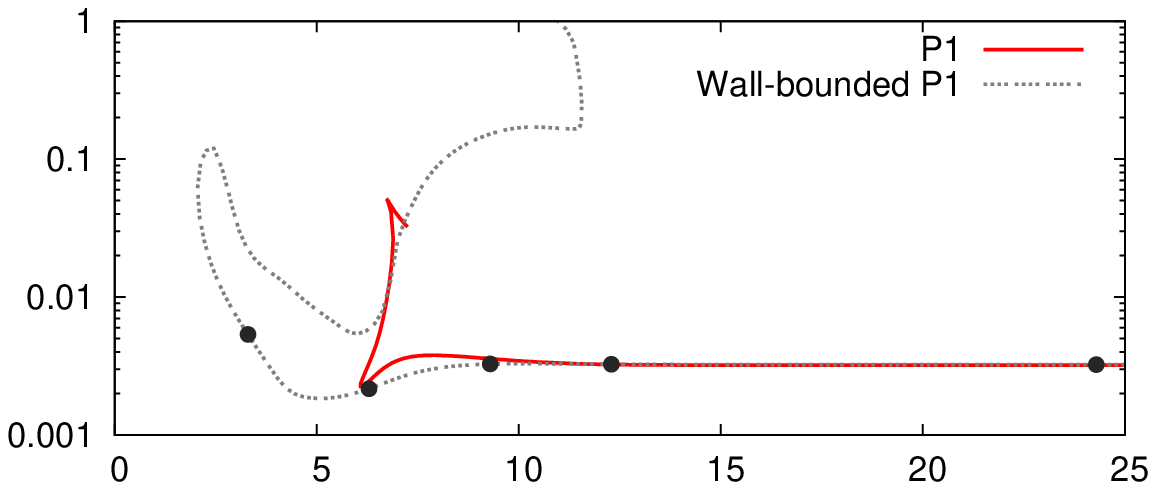}}}
\SetLabels 
(0.085*-0.175){\normalsize $L_z\approx24$} \\
\endSetLabels 
\leavevmode
\strut\AffixLabels{\includegraphics[height=14.75mm]{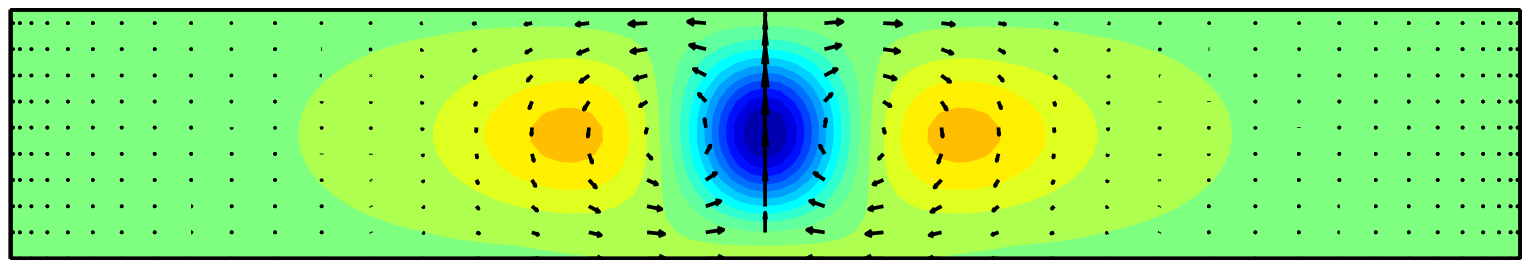}}\\
\SetLabels 
(-0.29*0.45){\normalsize $L_z\approx12$} \\
\endSetLabels 
\leavevmode
\strut\AffixLabels{\includegraphics[height=14.75mm]{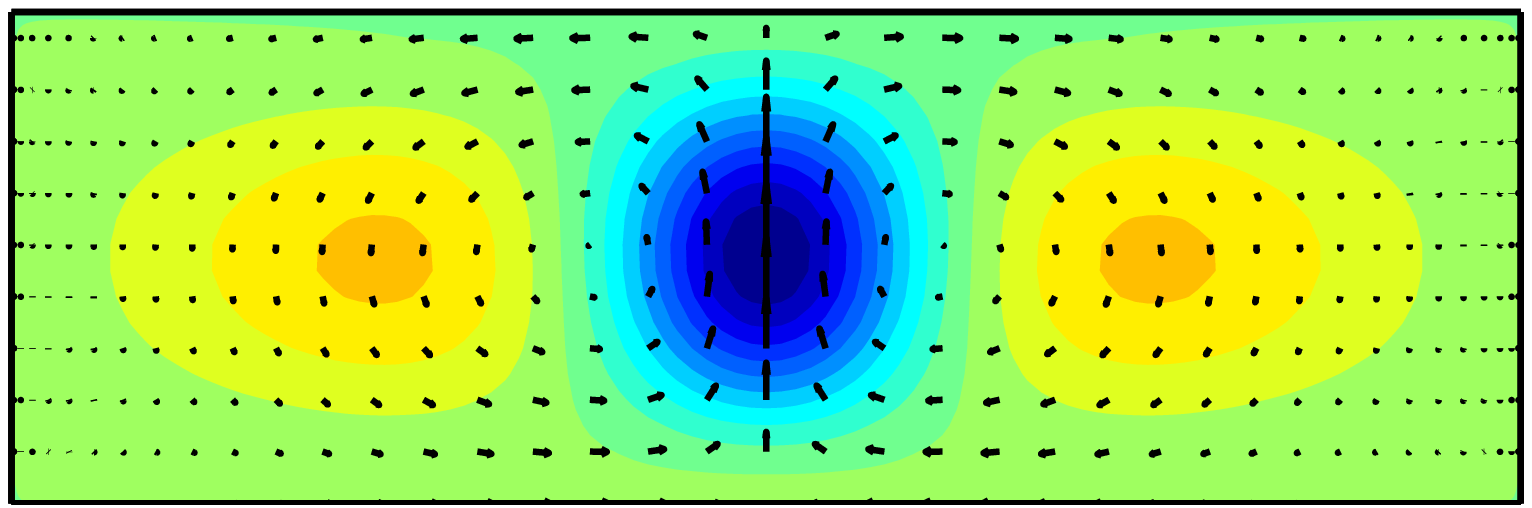}}\\
\SetLabels 
(-0.525*0.45){\normalsize $L_z\approx9.1$} \\
\endSetLabels 
\leavevmode
\strut\AffixLabels{\includegraphics[height=14.75mm]{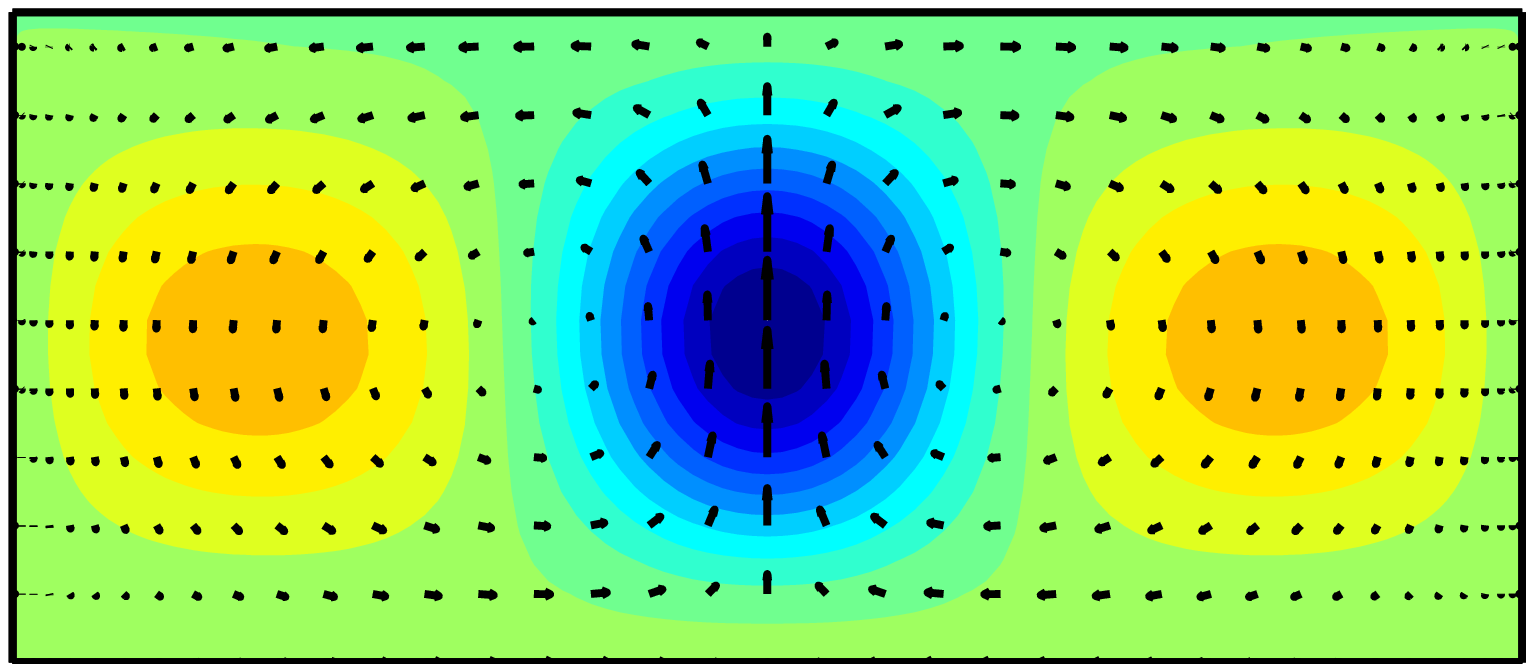}}\\
\SetLabels 
(-1.01*0.45){\normalsize $L_z\approx6.3$} \\
\endSetLabels 
\leavevmode
\strut\AffixLabels{\includegraphics[height=14.75mm]{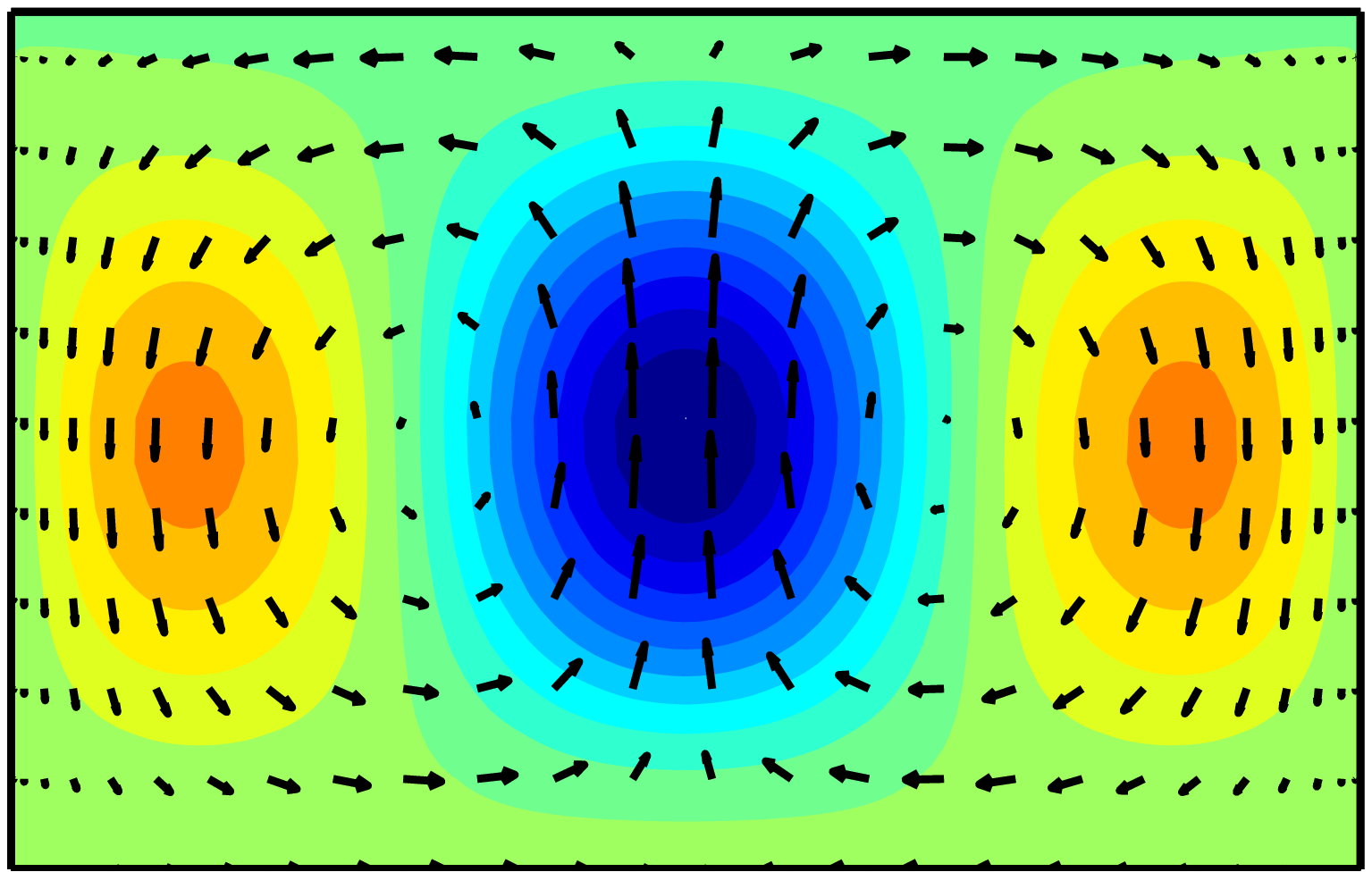}}\\
\SetLabels 
(-2.28*0.45){\normalsize $L_z\approx3.2$} \\
\endSetLabels 
\leavevmode
\strut\AffixLabels{\includegraphics[height=14.75mm]{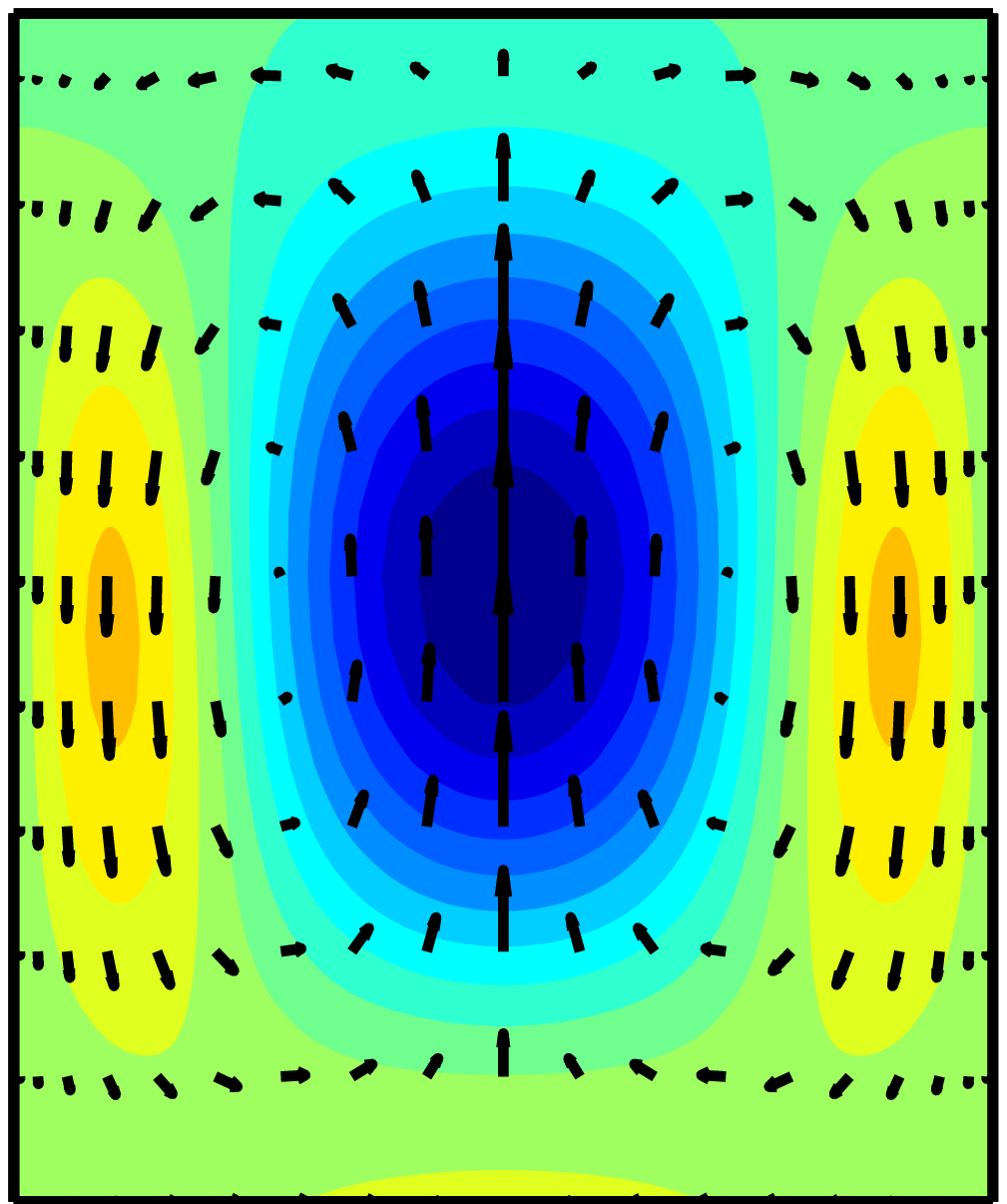}}
\end{center}
\caption{Continuation of localized solution $P1$ in $L_z$ for periodic
  boundary conditions (red) and spanwise-wall boundary conditions
  (grey dotted). Crosses denote locations of the solutions plotted
  below. Solutions are indistinguishable until $L_z\approx10$ when,
  compared to the periodic solution, the fast streaks (yellow) fail to
  grow to amplitude as domain width decreases.  }
\label{fig:P1Wall}
\end{figure}

\section{Conclusions}
\label{sec:conc}
In this work we have derived a 9-PDE extension to the 9-ODE Moehlis
model \citep{moehlis1} in order to study spanwise
localization. Despite tracking many steady modulational bifurcations
from 3 different global steady states, we found no tendency for
localization and therefore also no instance of homoclinic snaking.
Localized periodic orbits can, however, be found by
edge-tracking. When smoothly continued to smaller spanwise domains,
these look like global states as they fill the domain but are not
connected via any bifurcation to global states which only exist for a
finite range of domain sizes (e.g. figure \ref{fig:chantry}). This
behaviour is not dependent on the exact form of the spanwise boundary
conditions as it persists when no-slip sidewalls replace spanwise
periodicity.

The conclusions to be drawn from this model study are therefore
twofold: a) modulational instabilities leading to localisation from
global states are not generic at least when only considering steady
flows; and b) localized solutions can smoothly morph into global
solutions (and vice versa) without any need for a bifurcation in
between. This smooth local-to-global transition has not as yet been
observed in plane Couette flow but has been in fully resolved
computations in duct flow \cite{okino}. There, a square duct
travelling wave solution could be continued to a much wider domain
with little change to solution structure.

In terms of future work, the challenge still remains (surprisingly) to
demonstrate the generation of a localised state {\em from} a global
state by following a modulational instability in the Navier-Stokes
equations.  The few such bifurcations known have so far only be found
in `reverse' by tracking a localised solution back to its global
cousin \cite{schneider10localized, chantry2013}. The work reported
here has also highlighted the fact that localized solutions may not be
connected to {\em any} global states which only exist over a finite
range of domain sizes. Instead, they may be simply borne in a saddle
node bifurcation as the domain size increases.  Finally, we note that
the modelling strategy adopted here to develop a spanwise-extended
system could also equally be used to build a streamwise-extended model
of 1-space and 1-time PDEs. This smooth local-to-global transition may
well also exist for streamwise localisation too.

\vspace{1cm}
\noindent
Acknowledgements. Matthew Chantry is very grateful for the support of
EPSRC during his PhD.


\newpage
\appendix
\begin{widetext}
\section{Model Derivation}
\label{app:eqn}
The evolution equations for the modes listed in (\ref{modes}) are
derived using Fourier orthogonality in the $x$ and $y$ directions. To
simplify notation we introduce the following
$$A'_i \equiv \frac{{\partial} A_i }{\partial z}, \enskip
{\mathcal{D}_{\alpha}}^2 \equiv \left({\alpha}^2 - \frac{{\partial}^2
}{\partial z^2} \right), \enskip {\mathcal{D}_{\beta}}^2 \equiv
\left({\beta}^2 - \frac{{\partial}^2 }{\partial z^2} \right), \enskip
     {\mathcal{D}_{\alpha \beta}}^2 \equiv \left({\alpha}^2+\beta^2 -
     \frac{{\partial}^2 }{\partial z^2}\right).$$ 
To derive the equation for $A_1$ ($A_2$, $A_9$) we take the
$\mathbf{e_x}$ component of the Navier-Stokes, multiply this by the
$\mathbf{e_x}$ component of $A_1$ ($A_2$, $A_9$) and integrate over
$x$ and $y$,\\

$\int \int \left(\mathbf{u_1}\cdot\mathbf{e}_x\right) \left(\mathbf{NS}\cdot\mathbf{e}_x \right)\; dx dy \Rightarrow$
\begin{eqnarray}
\left(\frac{\partial }{\partial t} + \frac{1}{R} {\mathcal{D}_{\beta}}^2 \right) A_1 = 
 \frac{\alpha}{8} \left( A_4'' A_5 - A_4 A_5'' \right) 
 + \frac{\alpha}{8} \left( A_6 A_7'' - A_6'' A_7 \right)
  + \frac{\sqrt{2} {\beta}^2}{R}\nonumber\\
+ \frac{\beta}{4} A_6' {\mathcal{D}_{\alpha}}^2 A_8
 - \frac{\beta}{8} {\mathcal{D}_{\alpha}}^2 A_6 A_8'
 + \frac{\beta}{4} \left( 2 A_2 A_3' - A_2' A_3 \right),
\end{eqnarray}

$\int \int \left(\mathbf{u_2}\cdot\mathbf{e}_x\right) \left(\mathbf{NS}\cdot\mathbf{e}_x \right)\; dx dy \Rightarrow$
\begin{eqnarray}
\left(\frac{\partial }{\partial t} + \frac{1}{R} \left(\frac{4}{3}\beta^2 - \frac{{\partial}^2 }{\partial z^2} \right) \right) A_2 = 
 \frac{5 \alpha}{12} \left( A_4'' A_6 - A_4 A_6'' \right)
+ \frac{\alpha}{6} \left( A_5 A_7'' - A_5'' A_7 \right)\nonumber\\
- \beta \left( A_1 A_3' + A_3' A_9 \right) 
- \frac{\beta}{3} \left( A_1' A_3 + A_3 A_9' \right)\nonumber\\
- \frac{\beta}{6} \left( {\mathcal{D}_{\alpha}}^2 A_5 A_8' +3 A_5'  {\mathcal{D}_{\alpha}}^2 A_8 \right),
\end{eqnarray}

$\int \int \left(\mathbf{u_9}\cdot\mathbf{e}_x\right) \left(\mathbf{NS}\cdot\mathbf{e}_x \right)\; dx dy \Rightarrow$
\begin{eqnarray}
\left(\frac{\partial }{\partial t} + \frac{1}{R} \left({9 \beta}^2 - \frac{{\partial}^2 }{\partial z^2} \right) \right) A_9 = 
\frac{\beta}{4} \left( 2 A_2 A_3' - A_2' A_3 \right)
+\frac{\alpha}{8} \left( A_4'' A_5 - A_4 A_5'' \right)\nonumber\\
+\frac{\alpha}{8} \left( A_6 A_7'' - A_6'' A_7 \right)
+\frac{\beta}{8} \left( 2 A_6' {\mathcal{D}_{\alpha}}^2 A_8 - {\mathcal{D}_{\alpha}}^2 A_6 A_8' \right).
\end{eqnarray} 
To generate the equations for modes 3-7 we consider the curl of the Navier-Stokes,\\

$\int \int \left(\mathbf{u_3}\cdot\mathbf{e}_x\right) \left(\left(\nabla \times \mathbf{NS}\right)\cdot\mathbf{e}_x \right)\; dx dy \Rightarrow$
\begin{eqnarray}
\left(\frac{\partial }{\partial t} + \frac{1}{R} {\mathcal{D}_{\beta}}^2 \right) {\mathcal{D}_{\beta}}^2 A_3 = 
 \frac{3 {\alpha}}{8} \left( A_4 {\mathcal{D}_{\alpha}}^2 A_8 \right)''
 - \frac{{\alpha}^2 \beta}{4} \left( A_4 A_7 \right)'
 - \frac{{\alpha}^2 \beta}{4} \left( A_5 A_6 \right)'\nonumber\\
+ \frac{\alpha {\beta}^2}{8} \left( 2 A_4' A_8' - A_4 A_8'' \right)
 + \frac{3 {\alpha}^3 {\beta}^2}{8} A_4 A_8,
\end{eqnarray}

$\int \int \left(\mathbf{u_4}\cdot\mathbf{e}_y\right) \left(\left(\nabla \times \mathbf{NS}\right)\cdot\mathbf{e}_y \right)\; dx dy \Rightarrow$
\begin{eqnarray}
\left(\frac{\partial }{\partial t} + \frac{1}{R} \left( \frac{4}{3}\beta^2 +  {\mathcal{D}_{\alpha}}^2 \right)\right) {\mathcal{D}_{\alpha}}^2 A_4 = 
\alpha {\beta}^2 \left( A_3 A_8'' - A_3'' A_8 - {\alpha}^2 A_3 A_8 \right)\nonumber\\
+ \frac{\alpha}{3} \left( A_1 A_5'' - A_1'' A_5 - {\alpha}^2 A_1 A_5 \right)
+ \frac{5 \alpha}{6}\left( A_2 A_6'' - A_2'' A_6 - {\alpha}^2 A_2 A_6 \right)\nonumber\\
+ \frac{\alpha}{3} \left( A_5'' A_9 - A_5 A_9'' - {\alpha}^2 A_5 A_9 \right) 
- \frac{\beta}{3} \left( 4  A_3'  {\mathcal{D}_{\alpha}}^2 A_7 + A_3  {\mathcal{D}_{\alpha}}^2 A_7' - 3 A_3'' A_7' \right),
\end{eqnarray}

$\int \int \left(\mathbf{u_5}\cdot\mathbf{e_y}\right) \left(\left(\nabla \times \mathbf{NS}\right)\cdot\mathbf{e_y} \right)\; dx dy \Rightarrow$
\begin{eqnarray}
\left(\frac{\partial }{\partial t} + \frac{1}{R} {\mathcal{D}_{\alpha \beta}}^2 \right) {\mathcal{D}_{\alpha}}^2 A_5 = 
 \frac{\alpha}{4} \left( A_2'' A_7 - A_2 A_7'' + {\alpha}^2 A_2 A_7 \right)\nonumber\\
+ \frac{\alpha}{4} \left( A_4 A_9'' - A_4'' A_9 + {\alpha}^2 A_4 A_9 \right)
+ \frac{\alpha}{4} \left( A_1'' A_4 - A_1 A_4'' + {\alpha}^2 A_1 A_4 \right)\nonumber\\
+ \frac{\beta}{4} \left(  A_3' {\mathcal{D}_{\alpha}}^2 A_6 -  A_3 {\mathcal{D}_{\alpha}}^2 A_6' - 2 A_3'' A_6' \right) 
- \frac{\beta}{4} \left(   A_2' {\mathcal{D}_{\alpha}}^2 A_8 + 2   A_2 {\mathcal{D}_{\alpha}}^2 A_8'  + A_2'' A_8' \right) ,
\end{eqnarray}

$\int \int \left(\mathbf{u_6}\cdot\mathbf{e_y}\right) \left(\left(\nabla \times \mathbf{NS}\right)\cdot\mathbf{e_y} \right)\; dx dy \Rightarrow$
\begin{eqnarray}
\left(\frac{\partial }{\partial t} + \frac{1}{R}\left( \frac{4}{3}\beta^2 + {\mathcal{D}_{\alpha}}^2 \right)\right) {\mathcal{D}_{\alpha}}^2 A_6 = 
 \frac{\alpha}{3}  \left( A_1'' A_7 - A_1 A_7'' + {\alpha}^2 A_1 A_7 \right)\nonumber\\
+ \frac{\beta}{3} \left(3 A_1 {\mathcal{D}_{\alpha}}^2 A_8' + 4 A_1' {\mathcal{D}_{\alpha}}^2 A_8 - A_1'' A_8' \right)
+ \frac{\alpha}{3}  \left( A_7 A_9'' - A_7'' A_9 + {\alpha}^2 A_7 A_9 \right)\nonumber\\
+ \frac{\beta}{3} \left( 3{\mathcal{D}_{\alpha}}^2 A_8' A_9 + 4 {\mathcal{D}_{\alpha}}^2 A_8 A_9' - A_8'A_9''\right)
+ \frac{5\alpha}{6}  \left( A_2'' A_4 - A_2 A_4'' + {\alpha}^2 A_2 A_4 \right)\nonumber\\
- \frac{\beta}{3} \left( A_3 {\mathcal{D}_{\alpha}}^2 A_5' + 4 A_3' {\mathcal{D}_{\alpha}}^2 A_5 - 3 A_3'' A_5'\right),
\end{eqnarray}

$\int \int \left(\mathbf{u_7}\cdot\mathbf{e_y}\right) \left(\left(\nabla \times \mathbf{NS}\right)\cdot\mathbf{e_y} \right)\; dx dy \Rightarrow$
\begin{eqnarray}
\left(\frac{\partial }{\partial t} + \frac{1}{R} {\mathcal{D}_{\alpha \beta}}^2 \right) {\mathcal{D}_{\alpha}}^2 A_7 = 
 \frac{\alpha}{4} \left( A_1 A_6'' - A_1'' A_6 - {\alpha}^2 A_1 A_6 \right)\nonumber\\
 +\frac{\alpha}{4} \left( A_6'' A_9 - A_6 A_9'' - {\alpha}^2 A_6 A_9 \right)
 +\frac{\alpha}{4} \left( A_2 A_5'' - A_2'' A_5 - {\alpha}^2 A_2 A_5 \right)\nonumber\\
 +\frac{\alpha^2 \beta}{4} \left( A_3' A_4 - A_3 A_4' \right)
 -\frac{\beta}{2} \left( A_3' A_4'\right)' + \frac{\beta}{4} \left( A_3 A_4''\right)'.
\end{eqnarray}

Finally, we take the curl twice to produce the evolution equation for $A_8$,

$\int \int \left(\mathbf{u_8}\cdot\mathbf{e_y}\right) \left(\left(\nabla \times \left(\nabla \times \mathbf{NS}\right)\right)\cdot\mathbf{e_y} \right)\; dx dy \Rightarrow$
\begin{eqnarray}
\left(\frac{\partial }{\partial t} + \frac{1}{R} {\mathcal{D}_{\alpha \beta}}^2 \right) {\mathcal{D}_{\alpha \beta}}^2 {\mathcal{D}_{\alpha}}^2 A_8 = 
 \frac{\alpha \beta^2}{2} \left( 2 A_3'' A_4  - A_3' A_4' \right)
 -\frac{3 \alpha}{4} {\mathcal{D}_{\alpha \beta}}^2 \left( A_3'' A_4  \right) \nonumber\\
-\frac{\alpha^2 \beta}{2} \left( A_1' A_6 + A_2' A_5 + A_6 A_9' \right).
\end{eqnarray}
\end{widetext}
\bibliographystyle{jfm}
\bibliography{paper}

\end{document}